\documentclass[fleqn,usenatbib]{mnras}

\usepackage{newtxtext,newtxmath}

\usepackage[T1]{fontenc}

\DeclareRobustCommand{\VAN}[3]{#2}
\let\VANthebibliography\thebibliography
\def\thebibliography{\DeclareRobustCommand{\VAN}[3]{##3}\VANthebibliography}

\usepackage{graphicx}	
\usepackage{amsmath}	
\usepackage{booktabs}
\usepackage{graphicx,float, caption}

\usepackage{placeins}
\usepackage{natbib}
\usepackage{float}
\usepackage{textcomp} 
\usepackage{anyfontsize}

\title[M101, NGC~3938 and their significant others.]{
Uncovering the truth about M101, NGC~3938, and their significant others through radiative transfer}

\author[D. Pricopi et al.]{
D. Pricopi,$^{1}$
C. C. Popescu,$^{2}$\thanks{E-mail: CPopescu@uclan.ac.uk}
M. T. Rushton,$^{1,2}$
D. Murphy,$^{2}$
C. J. Inman,$^{2}$\thanks{E-mail: CJInman@uclan.ac.uk}
R. Toma$^{1}$
\\
$^{1}$Astronomical Institute of the Romanian Academy, Str. Cutitul de Argint 5, 040557, Bucharest, Romania\\
$^{2}$Jeremiah Horrocks Institute, University of Central Lancashire, Preston, PR1 2HE, UK\\
}

\date{Accepted XXX. Received YYY; in original form ZZZ}

\pubyear{2025}

\begin{document}


\defcitealias{PT11}{PT11}
\defcitealias{TP20}{TP20}
\defcitealias{DL14}{DL14}


\label{firstpage}
\pagerange{\pageref{firstpage}--\pageref{lastpage}}
\maketitle


\begin{abstract}
Solving the inverse problem in spiral galaxies, that allows the derivation of the spatial distribution of dust, gas and stars, together with their associated physical properties, directly from panchromatic imaging observations, is one of the main goals of this work. To this end we used radiative transfer models to decode the spatial and spectral distribution of the nearby face-on galaxies M101 and NGC~3938. In both cases we provide excellent fits to the surface-brightness distributions derived from GALEX, SDSS, 2MASS, 
Spitzer and Herschel imaging observations. Together with previous results from M33, NGC~628, M51 and the Milky Way, we obtain a small statistical sample of modelled nearby galaxies that we analyse in this work. We find that in all cases Milky Way-type dust with Draine-like optical properties provide consistent and successful solutions. We do not find any "submm excess", and no need for modified dust-grain properties. Intrinsic fundamental quantities like star-formation rates (SFR), specific SFR (sSFR), dust opacities and attenuations are derived as a function of position in the galaxy and overall trends are discussed. In the SFR surface density versus stellar mass surface density
space we find a structurally resolved relation (SRR) for the morphological components of our galaxies, that is steeper than the main
sequence (MS). Exception to this is for NGC 628, where the SRR is parallel to the MS. 
\end{abstract}




\section{Introduction} \label{sec:intro}

The advent of spatially resolved high quality multiwavelength photometric surveys of local galaxies spanning wavelengths from the space ultraviolet (UV) to the submillimetre, has opened the door to accurately characterise the complex interplay between the nature of the stellar structure, stellar populations, and the dusty interstellar medium (ISM) of star forming galaxies \citep{2011PASP..123.1347K}.
Studies using infrared space telescopes, starting with IRAS (Infrared Astronomical Satellite; \cite{1984ApJ...278L...1N}) and ISO (Infrared Space Observatory; \cite{1996A&A...315L..27K}), and continuing with Spitzer (Spitzer Space Telescope; \cite{2004ApJS..154....1W}) and Herschel (Herschel Space Observatory; \cite{2010A&A...518L...1P}), have revealed that while dust is usually a minor fraction of the baryonic mass of a galaxy, it is responsible for a significant fraction of the bolometric luminosity, averaging $\approx 25-30\%$ of the total luminosity for star forming late-type galaxies \citep{2002MNRAS.335L..41P,2018A&A...620A.112B}. Dust emission in galaxies is powered by the absorption of stellar photons by dust grains, thus the radiative energetics of  stellar populations and dust are linked by energy conservation. Because of this, understanding the radiative properties of galaxies inherently requires simultaneous consideration of UV/optical and far infrared (FIR)/submillimetre emission.

Dust in star forming galaxies, as with stars, is spatially distributed through their extent, often leading to significant and varying optical depth along observational lines-of-sight, depending strongly on both intrinsic (e.g. dust properties; spatial distributions of stars relative to the dust) as well as extrinsic galaxy properties (e.g. inclination) (Kylafis \& Misiriotis \citeyear{D_Kylafis_2006}). Utilizing UV and optical images of dusty galaxies to determine the intrinsic geometrical distribution of stellar populations from observations must therefore account for the altering effects of both scattering and absorption of starlight by dust. Disentangling the underlying physical properties of late-type galaxies from observations, are thus best suited for radiative transfer (RT) models, that self-consistently include the effects of both scattering, absorption, and re-radiation of dust with realistic assumptions regarding the spatial distribution of stars and dust. Furthermore, as the spectral energy distributions (SEDs) of dusty galaxies evince energetically significant, complex, and varying emission (as a function wavelength) from the UV to submillimetre (from stars, dust, or both), models and the observations that constrain them, must similarly span a wide range of wavelengths. 

To address these considerations, RT models for star forming galaxies have been developed over the past decades: \cite{1998A&A...331..894X, 2000A&A...362..138P,2008A&A...490..461B,2011ApJS..196...22B,PT11,2015A&C....12...33B}. Recently a lot of work has been devoted to model nearby face-on spiral galaxies. The SKIRT code \citep{2015A&C....12...33B} has been used for non-axi-symmetric modelling: \cite{2014A&A...571A..69D} and \cite{2020A&A...643A..90N} for
M51, \cite{2017A&A...599A..64V} for M31, \cite{2019MNRAS.487.2753W} for M33,
\cite{2020A&A...637A..24V} for M81, \cite{2020A&A...638A.150V} for NGC 1068,
and \cite{2020IAUS..341...65N} for 4 barred galaxies. The codes from \cite{PT11} were used for axi-symmetric RT modelling: \cite{TP20} for M33, \cite{2022MNRAS.514..113R} for NGC~628, and \cite{2023MNRAS.526..118I} for M51.
 
Non-axi-symmetric models are, in principle, ideal to capture the detailed geometry seen face-on. However, due to computational limitations, they have been implemented to only fit the integrated SED, with the geometry being fixed. The axi-symmetric RT models have been implemented to also fit the geometry of stars and dust, albeit with the assumption of axi-symmetry. As mentioned before, only three face-on galaxies have been modelled with axi-symmetric codes. The main goal of this paper is to further the application of these models to the local galaxies M101 and NGC~3938. Another goal is to construct a small sample of face-on galaxies, all modelled using the same RT codes and modelling procedures, to allow the characterisation of fundamental intrinsic properties of nearby star forming galaxies, including their spatial distributions of star-formation rates (SFRs), specific star-formation rates (sSFRs), stellar masses, dust masses, attenuation laws, fractions of young and old stellar populations in heating the dust.

The  multi-wavelength axi-symmetric RT model of Popescu et al. \citeyear{PT11} (hereafter \citetalias{PT11}) was thus used for modelling M101 and NGC~3938. 
Our model is three-dimensional in nature, but assumes an axi-symmetric geometry, a compromise necessitated, as mentioned before, for computational feasibility. The output of our model consists of galaxy images from starlight and dust, at the same wavelengths and sky projections as the observed galaxies. Our model details were given by \citetalias{PT11} who first applied it to model the edge-on spiral galaxy NGC~891. Additional refinements needed to model nearby face-on galaxies were presented by Thirlwall et al. \citeyear{TP20} (hereafter \citetalias{TP20}), in their  application to model the spiral M33 (see also the application to NGC~628 by Rushton et al. \citeyear{2022MNRAS.514..113R}).
A few salient details regarding the model are discussed below in Sec. \ref{sec:model}, however a complete description is beyond the scope of this paper and the reader is referred to the aforementioned references for full details.

As mentioned before, the main goal of this paper is to further the application or our RT model by applying it to the local galaxies M101 (NGC 5457) and NGC 3938. These galaxies are both late type, nearly face-on local spirals. The face-on nature of these galaxies has the benefit of reducing geometric uncertainties associated with inclination effects (although our RT model is inherently able to account for inclination effects). NGC 3938 and M101 are classified as Hubble numerical T stages 5 and 6, respectively, both equating to the same standard Hubble classification of Sc (de Vaucouleurs et al. \citeyear{1991rc3..book.....D}). Both NGC 3938 and M101 are known to have central bulges (Bottema \citeyear{1988A&A...197..105B}; Okamura et al. \citeyear{1976PASJ...28..329O}). Of the two galaxies, M101 is particularly well-studied. Also motivating the application of our model, is the fact that high quality multi-wavelength photometric imaging data from the near-UV to the sub-millimetre wavelengths is available for M101 and NGC 3938 with both being spatially well-resolved. The presence of spiral arms in M101 and NGC 3938 obviously indicates a degree of non-axial symmetry to these galaxies. This fact does not, however, impede the use of our axi-symmetric model. 
\citetalias{PT11} demonstrated that by taking the spiral arm structure into account (together with all other assumptions of our model) SEDs are predicted to differ negligibly from SEDs  predicted assuming axisymmetry. Other works have found similar results (see \citetalias{PT11} and references therein for a full discussion, as well as \cite{2014MNRAS.438.3137N}).

The aforementioned characteristics of M101 and NGC 3938 thus make them desirable candidates for the application of our model. Despite the similarities between these two galaxies, prior studies have revealed some important differences. For instance, it is known that M101 has an HI disk extending far beyond its optical disk, which may be interacting with the less massive spirals NGC 5477 and NGC 5474 (evident in the extended field of M101) as well as with other discovered massive HI clouds not associated with known galaxies (Rownd et al. \citeyear{1994AJ....108.1638R}; Mihos et al. \citeyear{2013ApJ...762...82M}; Xu et al. \ \citeyear{2021ApJ...922...53X}). Also evident on the very deep optical images of M101 of Mihos et al. \citeyear{2013ApJ...762...82M} are a faint optical "plume" and "spur", stellar emission features extending   northeastward and eastward from the main bright optical disk of the galaxy, respectively. These features are of uncertain origin, but are possibly relics of past interactions with the other galaxies seen in close projection to M101 (Mihos et al. \citeyear{2013ApJ...762...82M}). The dynamical simulation of \cite{2022ApJ...933L..33L} showed that the extended NE plume and the sharp western edge of M101\textquotesingle s disk could be explained by a relatively close passage of the companion galaxy NGC 5474 through M101\textquotesingle s outer disk, approximately 200 Myr ago.

As with NGC 3938, M101 is a member of a galaxy group consisting of up to 15 possible members, the M101 Group, with M101 as its brightest member (Fouque et al. \citeyear{1992A&AS...93..211F}). NGC 3938 appears more symmetric in appearance than M101, lacking close ($<100$\,kpc) companions and other clear evidence of external interactions (Jim{\'e}nez-Vicente et al. \citeyear{1999A&A...342..417J}). Attempting to better understand the observational differences and similarities between M101 and NGC 3938 and how they are reflected (or not) in the physical properties of the underlying stellar and dust populations derived from our model is a goal of our study. One aspect we seek to understand in particular is if the derived physical properties bear any relationship to the nearby external environment of these galaxies, one seemingly isolated from the effects of nearby companions and intergalactic gas structures (NGC 3938), the other not (M101). Another goal of our study is to further test the predictive ability of our RT model and to compare its physical predictions for M101 and NGC 3938 to independent determinations from other studies and as well to compare our model's predictions to those of similar spiral galaxies to which we have applied it.

Bosma et al. \citeyear{1981A&A....93..106B} reported an inclination of $18^{\circ}$ determined from HI observations for M101. As the nearest low inclination spiral, M101 has been a popular target for studies of external galaxies made across a wide range of wavelengths. Fortunately, M101 also has a recent and reliable distance estimate of $6.71^{+0.14}_{-0.13}$ Mpc, obtained from its well-studied Cepheid variable population which serve as an important component of the cosmological distance ladder (Riess et al. \citeyear{2016ApJ...826...56R}). 

Jim{\'e}nez-Vicente et al. \citeyear{1999A&A...342..417J} determined an inclination of $14^{\circ}$ for NGC 3938 based on H${\alpha}$ line observations. This galaxy is also a commonly studied object of nearby galaxy surveys, but somewhat less so than M101. The distance to NGC 3938 is less accurately known than that of M101 as Cepheid variable or type Ia supernova based measurements are unavailable. Here we will adopt the distance determination of $17.9^{+1.2}_{-1.0}$ Mpc to NGC 3938 of Poznanski et al. \citeyear{2009ApJ...694.1067P} based on the type II-P supernova 2005ay, which these authors employ as a standard candle, This choice is also consistent with the estimated distance range of 17.0 to 22.4 Mpc adopted by Van Dyk et al. \citeyear{2018ApJ...860...90V} in identifying a candidate progenitor to the NGC 3938 type Ic supernova SN2017ein. We note that NGC 3938 is a member of the Ursa Major galaxy cluster, a fairly sparse cluster of mostly HI rich spiral galaxies at 18.6 Mpc as determined by the Tully-Fisher relationship,  consistent with the aforementioned distance estimates (McDonald et al. \citeyear{2009MNRAS.393..628M}; Tully \& Pierce \citeyear{2000ApJ...533..744T}). 

The paper is organised as follows: In Sec.~\ref{sec:observations} we describe the panchromatic images of M101 and NGC~3938 and in Sec.~\ref{sec:SB_prof} we describe the data reduction procedure. The model is described in Sec.~\ref{sec:model}. Results regarding the main characteristics of our model fits and of the intrinsic volume stellar emissivity functions and of the dust density functions, the derived SEDs and corresponding intrinsic properties are given in Sec.~\ref{sec:Results}. In Sec.~\ref{sec:discussion} we discuss the physical implications of our results for M101 and NGC~3938. For comparative purposes, we also analyse results coming from a small statistical sample of nearby galaxies that we assemble to contain, besides M101 and NGC~3938, galaxies modelled in previous works: NGC~628, M33, M51 and the Milky Way. We give our conclusions in Sec.~\ref{sec:conclusions}.

\section{Observations}
\label{sec:observations}

Panchromatic images of M101 and NGC 3938 were compiled from archival sources, 
yielding similar wavelength coverage for both galaxies, from the UV to sub-millimetre wavelengths. The named wavebands and equivalent wavelength (in $\micron$) used for this study are listed in columns one and two of Table \ref{tab:obs_flux}, respectively. The sources for our image data are
described in the subsections below.

\subsection{GALEX}
\label{galex}

The Galaxy Evolution Explorer (GALEX; Martin et al. \citeyear{2005ApJ...619L...1M}) conducted an all-sky survey as well as a deeper imaging survey known as the Nearby Galaxy Survey (NGS; Gil de Paz et al. \citeyear{2007ApJS..173..185G}) in the far ultraviolet (FUV, $0.15\,\mu$m) and near UV (NUV; $0.23\,\mu$m). We employ the FUV and NUV images from the NGS to model M101. As NGC 3938 was absent from the NGS, and the FUV GALEX all-sky survey was found to be too noisy for our analysis, we only employed the all-sky survey NUV imaging of NGC 3938 to fit our model. We used the factors given in Morrissey et al. \citeyear{2007ApJS..173..682M} to convert GALEX image data units from counts per second to flux units.  

\subsection{SDSS}

The optical data used to fit our model originated from Sloan Digital Sky Survey (SDSS) images of M101 and NGC 3938 made in the ugriz bands. Because of the significant extent of M101 on the sky ($\sim 30$ arcsec at visual wavelengths), it was necessary to create image mosaics for each SDSS band. This was accomplished with the application SWarp (Bertin et al. \citeyear{2002ASPC..281..228B}), ensuring the presence of the entire galaxy on analyzed images, necessary to create the surface brightness (SB) profiles  to fit our model (see Sec. \ref{sec:SB_prof} below). NGC 3938 is less extended on the sky allowing the entire galaxy to be contained within a single SDSS field. 

\subsection{2MASS}

To provide coverage of the near-infrared (NIR) bands, we downloaded and analyzed  J, H, and K band images from the 2MASS Large Galaxy Atlas (Jarrett et al. \citeyear{2003AJ....125..525J}), itself based on the Two Micron All Sky Survey (2MASS). Flux units were obtained from pixel data values using the photometric zero points supplied in image FITS Headers.      

\begin{table*}
\caption{Observed spatially integrated flux densities in Jy for NGC 3938 and M101 for all modeled wavebands. Our results are shown along side those of Dale et al. \citeyear{2017ApJ...837...90D} for comparison. 
The flux densities in this table are purely observed (not corrected for effects of MW foreground extinction.)}
  \begin{tabular}{ l | c | c | c | c  | c}
  \toprule
 $\lambda_{\rm eff}$ ($\mu$m) &  Band name & \multicolumn{4}{c}{Flux (Jy)}   \\
                     &   &    \multicolumn{2}{c}{NGC 3938}       &      \multicolumn{2}{c}{M101}   \\
\cmidrule(lr){3-4}\cmidrule(lr){5-6}
             &                  &      This paper     &  Dale et al. (2017)  &         This paper     &  Dale et al. (2017) \\
\midrule
      0.15   &    GALEX FUV     &              --          &     --               &       $0.3833\pm0.043$  & $0.346\pm0.054$  \\
     0.22    &    GALEX NUV     &    $0.04599\pm0.0046$     &  $0.0311\pm0.0043$   &       $0.5488\pm0.060$  & $0.387\pm0.059$   \\
     0.36    &    SDSS u        &   $0.08222\pm0.0016$     &  $0.0724\pm0.0015$   &       $0.7966\pm0.092$  & $0.755\pm0.015$   \\
     0.48    &    SDSS g        &   $0.21889\pm0.0044$     &  $0.203\pm0.004$     &       $1.8897\pm0.053$  & $1.89\pm0.03$    \\
     0.62    &    SDSS r        &   $0.33230\pm0.0066$     &  $0.314\pm0.003$     &       $2.800\pm0.074$   & $2.75\pm0.05$   \\
     0.77    &    SDSS i        &   $0.42099\pm0.0084$     &  $0.407\pm0.008$     &       $4.605\pm0.119$   & $3.52\pm0.07$   \\
     0.91    &    SDSS z        &   $0.48718\pm0.0097$    &  $0.464\pm0.009$     &       $3.699\pm0.101$   & $3.86\pm0.07$   \\  
      1.26   &    2MASS J       &   $0.65760\pm0.0132$    &  $0.625\pm0.031$     &       $5.705\pm0.146$   & $4.35\pm0.21$   \\
      1.65   &    2MASS H       &   $0.66560\pm0.0133$    &  $0.570\pm0.029$     &       $5.771\pm0.129$   & $5.01\pm0.25$   \\
      2.20   &    2MASS K       &   $0.52400\pm0.0105$    &  $0.531\pm0.027$     &       $4.129\pm0.093$   & $4.40\pm0.22$   \\
      3.60   &    IRAC 1        &   $0.36810\pm0.0074$    &  $0.320\pm0.043$     &       $2.549\pm0.060$   & $2.81\pm0.38$    \\
      4.50   &    IRAC 2        &   $0.18364\pm0.00367$   &  $0.211\pm0.029$     &       $1.655\pm0.060$   & $1.89\pm0.25$          \\
      5.67   &    IRAC 3        &    $0.48157\pm0.0096$    &  $0.410\pm0.052$     &       $4.219\pm0.091$   & $3.38\pm0.42$   \\
      8.00   &    IRAC 4        &   $1.35090\pm0.0270$    &  $0.98\pm0.122$      &       $10.23\pm0.243$   & $7.61\pm0.94$  \\
      24     &    MIPS 24       &   $1.06857\pm0.0214$     &  $1.08\pm0.04$       &       $10.64\pm0.255$   & $10.5\pm0.4$   \\
      70     &    PACS 70       &   $15.4487\pm0.7724$    &  $16.4\pm0.8$        &       $110.3\pm5.740$   & $135\pm6$      \\
      100    &    PACS 100      &   $28.6395\pm1.4320$     &  $30.7\pm1.5$        &       $235.8\pm12.91$   & $262\pm13$    \\
      160    &    PACS 160      &   $40.3950\pm2.0198$     &  $39.2\pm1.9$        &       $340.0\pm33.82$   & $336\pm16$    \\
      250    &    SPIRE 250     &   $23.0055\pm1.1503$     &  $21.9\pm1.5$        &       $211.1\pm10.77$   & $193\pm13$     \\
      350    &    SPIRE 350     &   $10.2644\pm0.5130$     &  $9.83\pm0.69$       &       $110.4\pm5.720$   & $94.2\pm6.6$   \\
      500    &    SPIRE 500     &   $3.89150\pm0.1946$    &  $3.65\pm0.26$       &       $43.86\pm3.03$    & $38.8\pm2.7$  \\
\bottomrule
  \end{tabular}
  \label{tab:obs_flux} 
  \end{table*}  

\subsection{Spitzer} \label{Spitzer_sec}

The Spitzer Space Telescope was a 0.85 m telescope, which conducted infrared imaging with two onboard instruments: the Infrared Array Camera (IRAC; Fazio et al. \citeyear{2004ApJS..154...10F}) and the Multiband Imaging Photometer for Spitzer (MIPS;  Rieke et al. \citeyear{2004ApJS..154...25R}). M101 and NGC 3938 were observed as part of the Spitzer Infrared Nearby Galaxies Survey (SINGS; Kennicutt et al. \citeyear{2003PASP..115..928K}), with images taken at 3.6 $\micron$, 4.5 $\micron$, 5.7 $\micron$ and 8.0 $\micron$ (IRAC), and 24 $\micron$ (MIPS), all of which were used in our study. SINGS also imaged M101 and NGC 3938 at 160 $\micron$ with MIPS, but we excluded this data from our analysis, preferring to use the Herschel Space Observatory image (Sec. \ref{Herschel}) observed at similar effective wavelength, but with superior sensitivity and spatial resolution.

\subsection{Herschel}
\label{Herschel}

The Herschel Space Observatory was a 3.5 m telescope dedicated to FIR and submillimetre-wave observations. M101 and NGC 3938 were observed as part of the Key Insights of Nearby Galaxies: A Far Infrared Survey with Herschel program (KINGFISH; Kennicutt et al. \citeyear{2011PASP..123.1347K}). For our study we used KINGFISH images obtained by the Photodetecting Array Camera and Spectrometer (PACS; Poglitsch et al. \citeyear{2010A&A...518L...2P}), and the Spectral and Photometric Imaging Receiver (SPIRE; Griffin et al. \citeyear{2010A&A...518L...3G}). These instruments provided image data at 70 $\micron$, 100 $\micron$, and 160$\micron$ (PACS), and at 250 $\micron$, 350 $\micron$, and 500 $\micron$ (SPIRE).

\section{Data reduction} \label{sec:SB_prof}

To prepare the observed images for comparison with our model, we first de-reddened the images for the effects of MW foreground extinction. While M101 and NGC 3938 are located on the sky at high galactic latitudes ($+59.8^{\circ}$ for M101 and $+69.3^{\circ}$ for NGC 3938), galactic extinction can still be a significant source of error if not corrected, particularly at the shortest wavelengths ($\lessapprox0.48\,\mu$m). We therefore de-reddened the UV and optical data, adopting colour excesses derived from the reddening maps of Schlafly \&  Finkbeiner \citeyear{2011ApJ...737..103S}: $E(B-V)=0.02$ (NGC 3938) and 0.007 (M101), and using the extinction curve of Fitzpatrick (\citeyear{1999PASP..111...63F}).

For use in fitting the free parameters of our axially symmetric RT model, a first step was to obtain azimuthally averaged SB profiles from the NGC~3938 and M101 data images, obtained by calculating the flux in a series of nested, two-pixel wide elliptical annuli around the centres of these galaxies. Before creating SB profiles this way, it was necessary to determine suitable axial ratios (${\rm b/a}$) and lines-of-nodes position angles (PAs) on the sky for the elliptical annuli. The axial ratios were found by fitting the SDSS and 2MASS data images with the 2D decomposition code GALFIT (Peng et al. \citeyear{2002AJ....124..266P}), using two component models, including an exponential disk and de Vaucouleurs bulge. The best compromise fitted values obtained, and those adopted, were ${\rm b/a}=0.96$ and ${\rm b/a}=0.97$, for M101 and NGC 3938, respectively. These determinations are consistent with the known low inclinations obtained independently by Jim{\'e}nez-Vicente et al. \citeyear{1999A&A...342..417J} (cos($18^{\circ}$)) = 0.95) and Bosma et al. \citeyear{1981A&A....93..106B} (cos($14^{\circ}$) = 0.97) (Sec. \ref{sec:intro}) for M101 and NGC 3938, respectively. At such small axial ratios, azimuthally averaged surface brightness profiles are largely insensitive to adopted PAs. The PA values were also determined by applying GALFIT, to the UV-optical-NIR images, obtaining values of $6^{\circ}$ and $38^{\circ}$ for NGC 3938 and M101, respectively. We note that the PA for NGC 3938 was poorly constrained by GALFIT, with values ranging from $6^{\circ}$ to $63^{\circ}$, found from fitting different NGC 3938 images, depending on wavelength. The small inclination of NGC 3938 makes this a difficult parameter to constrain, but as expected, the actual SB profiles calculated from the NGC 3938 images for use in fitting our model, were found to vary little as a function of the chosen PA within the above range. We emphasize that our PA values follow the usual convention of being measured from zero at due north, increasing in value counter-clockwise to the east. The SB, $\overline{I_\nu}$, of an annulus is defined as $\int I_{\nu} d\Omega / \int d\Omega$, the average of specific intensity $I_{\nu}$ over the annulus solid angle. To determine fitted RT model parameters, SB profiles of both model generated images and data images were calculated identically, using the axial ratios and PA values given above, and fitted by varying model parameters.

The errors in the observational SB profiles are derived in accordance with Appendix A of \citetalias{TP20} and account for four sources of error: calibration, background fluctuation, Poisson noise, and configuration noise. The last of these four errors measures the inherent uncertainty of the axi-symmetric approximation.

The spatially integrated flux densities of M101 and NGC 3938 at each sampled wavelength were derived from the curve-of-growth, a cumulative sum of the total flux within the elliptical annuli plotted as a function of distance from a galaxy centre. The sources of error are a sum-in-quadrature of calibration uncertainties and background fluctuations. The results are given in Table~\ref{tab:obs_flux}, where we compare them to the results of Dale et al. \citeyear{2017ApJ...837...90D}, who compiled photometry from 34 bands for the SINGS/KINGFISH Hershel/Spitzer sample of local galaxies.

Foreground stars contaminate the galaxy emission, producing spikes in the azimuthally averaged surface brightness profiles, and can also contribute to spatially integrated flux density uncertainties. Therefore, we endeavoured to remove foreground stars from the data images with a program that masks point sources on the observational images. Each masked image was carefully inspected to ensure that galaxy H\,{\sc ii} regions, which may appear as point sources, were left unmasked after this process. To help identification  additional colour information was used. In particular we looked at the 24\,$\mu$m images to identify coincidences between masked point sources in the optical/NIR and sources of localised hot dust emission originating from HII regions. Furthermore, we checked with catalogues of known star forming regions, where information was available. Manual masking for bad pixels was also necessary for some of the IRAC and PACS data images.

The physical resolution of our model is 50 pc. In cases where the physical resolution of the model was better than the FWHM resolution of the observational data images, we convolved the model  to the data resolution employing convolution kernel\textquotesingle s for GALEX, Spitzer, Herschel from Aniano et al. \citeyear{2011PASP..123.1218A}, and appropriately sized Gaussian kernels to match the 2MASS and SDSS images.
We note here that the fixed 2-pixel wide aperture used in producing the azimuthally averaged surface-brightness profiles will sample different physical scales at different wavelengths. Nonetheless, it is important to sample the data at their original resolution to get the best constraints on the model. Because of this we did not try to fit the model by sampling the profiles at a fixed physical size (of the lowest resolution data - the Herschel 500\,${\mu}$m). We performed, however,  some tests whereby the (already) best fitted model (fitted at the original resolution of the data) and the observed surface brightness profiles were sampled at a fixed physical size (see Fig.~\ref{fig:aperture_test}). Overall the main characteristics of the profiles remain the same, although some features are less prominent (see further discussion in Appendix~\ref{app:photometry}.)

\section{The model}
\label{sec:model}

\subsection{Model description}
\label{sec:model_descrp}

Our model follows the generic formalism described in \citetalias{PT11}, with the additional modifications described in \citetalias{TP20}. Briefly, the model assumes a galaxy structure consisting of a few sets of concentric disks of stars and dust with stellar volume emissivities (typical units W Hz$^{-1}$ pc$^{-3}$) and dust densities (typical units M$_{\odot}$ pc$^{-3}$) described by a modified analytic exponential function. A superposed $\textit{bulge}$ may also be part of the model with a stellar volume emissivity described by a S\'ersic function. We refer to the bulge and a set of disks as a \textit{morphological component} and galaxies may have more than one morphological component. Dust grains in the dust disks of morphological components can scatter and absorb both UV and optical photons, and re-radiate absorbed photons (primarily at longer wavelengths). As our RT model accounts fully for scattering, grains in the dust disks of any morphological component can interact with stellar photons originating from any disk  in any  morphological component. The simplest (initial) form of the model contains two morphological components, a bulge and a disk morphological component. The bulge is a dust-free stellar spheroid with an evolved stellar population and is located in the centre of a galaxy. A disk morphological component has a vertical stratification, containing a thinner and thicker component  of stellar and dust disks, as follows: 1) A disk comprised of a thin layer of young population stars (the \textit{thin stellar disk}); 2) A disk comprised of a thin layer of dust (the \textit{thin dust disk}). The thin dust disk is physically associated with the thin stellar disk; 3) A disk of evolved stellar population stars with a vertical scale height greater than the thin stellar disk (the \textit{stellar disk}); 4) A dust disk (the \textit{dust disk}) to account for dust associated with neutral hydrogen (HI) and of greater vertical scale height than the thin dust disk.

We have also included a clumpy component in our model to account for mid-IR emission (MIR) due to reprocessed UV photons from the youngest massive stars from stellar birth clouds. The model assumes that photons from these stars are totally absorbed by the dust of their dense, clumpy, progenitor molecular cores. 
The clumpy component assumes star/cloud clumps are distributed throughout galaxy thin disks with the same geometric parameters describing the thin stellar and thin dust disk density distributions. The radiative contribution of clumpy component galaxy emission is characterised by the model free parameter $F$ (\citetalias{PT11}), the fraction of UV light absorbed locally by progenitor molecular cloud clumps.   

Our basic model can be adapted, as necessary, to each individual galaxy, increasing its complexity in terms of morphological components as needed to achieve a good fit to observational data. Thus a galaxy may consist of more than one morphological component, each with its  component disks (vertical stratification). For example, a particular galaxy could require inner and outer morphological disk components, each with its own dust disk, thin dust disk, stellar disk, and thin stellar disk. This approach was found necessary to model the spirals M33 (\citetalias{TP20}), NGC 628 (Rushton et al. \citeyear{2022MNRAS.514..113R}), and M51 (Inman et al. \citeyear{2023MNRAS.526..118I}), as well as the Milky Way (Natale et al. \citeyear{2022MNRAS.509.2339N}). 

As the vertical scale heights of face-on galaxies cannot be uniquely constrained by our model, vertical scale heights are set to constant values to reflect typical scale heights of edge-on disk galaxies (Xilouris et al. \citeyear{1999A&A...344..868X}), including the MW (Natale et al. \citeyear{2022MNRAS.509.2339N}). Tests made with our model have shown that resulting fit parameters are relatively insensitive to the actual chosen values for scale heights (\citetalias{TP20}), so long as the relative ratios between stellar and dust scale heights follow adopted trends, first derived in Xilouris et al. \citeyear{1999A&A...344..868X}. The thin dust disk and thin stellar disk vertical scale heights were set to the same value, 90 pc, the typical height of the molecular layer of a MW type galaxy.

To model a galaxy, 
SB profiles created from the model images generated by the RT codes are fit to azimuthally averaged SB profiles measured from observed images, at every modelled waveband. Two RT codes were used in this analysis; the principle code is described in PT11, with certain ancillary calculations and checks provided by code DART-RAY (Natale et al. \citeyear{2014MNRAS.438.3137N}, \citeyear{2015MNRAS.449..243N}, \citeyear{2017A&A...607A.125N}). Our codes employ a ray-tracing algorithm accounting for both the absorption and anisotropic scattering of stellar light by dust grains of various sizes and chemical composition. The modeled grains are of silicate, graphite, and PAH composition, with a grain-size distribution and optical constants adopted from Weingartner \& Draine \citeyear{2001ApJ...548..296W}, Draine \& Lee \citeyear{1984ApJ...285...89D}, and Draine \& Li \citeyear{2007ApJ...657..810D}. We refer the reader to the aforementioned references for a detailed description of the dust models. 

The parameterized functional forms describing the stellar volume emissivities and dust disk density distributions of morphological component disks are given by
\begin{equation}
\label{eq:model}
w(R,z) = \begin{cases}
    {\displaystyle
     =0
    \hspace{3.75cm}{\rm if} \hspace{0.2cm} R < R_{{\rm tin}}
    }\\ \\
    {\displaystyle
        = A_0 \left[ \frac{R}{R_{\rm in}}\left( 1 - \chi \right) + \chi \right]} \\
        \hspace{.5cm}
        \times \hspace{.15cm} \exp{\left(-\frac{R_{{\rm in,}}}{h_0} \right)}
        \exp{\left(-\frac{z}{z_0}\right)}

    {\displaystyle
        \hspace{0.4cm}{\rm if} \hspace{0.1cm} R_{{\rm tin}} \leq R < R_{{\rm in}}
    }\\

    {\displaystyle
        = A_0 \exp{\left(-\frac{R}{h_0}\right)}
        \exp{\left(-\frac{z}{z_0}\right)}
         \hspace{.6cm} {\rm if} \hspace{0.1cm} R_{{\rm in}} \leq R \leq R_{{\rm t}}
    }\\
     {\displaystyle
     =0
    \hspace{3.75cm}{\rm if} \hspace{0.2cm} R > R_{{\rm t}}
    }\\
\end{cases}
\end{equation}
with
\begin{equation}
    \label{eq:chi}
\hspace{2cm}    \chi = \frac{w(0,z)}{w(R_{{\rm in}},z)}.
\end{equation}

\noindent
Here $w(R,z)$ represents the fitted stellar volume emissivity (e.g. units W Hz$^{-1}$ pc$^{-3}$) density for stellar disks, or the dust density for dust disks (e.g. units M$_{\odot}$ pc$^{-3}$). These distribution functions are functions of both radial coordinate $R$ from the galaxy centre, and height $z$ above the galaxy plane. In simplest terms, the functional character of these distributions is to model a physical scenario where disk stellar volume emissivity and dust density do not always increase exponentially up to the centre, but have \lq holes\rq\,or flattened emissivities in their centres.
These distributions thus represent a modified version of the classical exponential disk. The quantity $A_0$, an "amplitude" factor, is the stellar volume emissivity at the galaxy centre for stellar volume emissivity distributions or the central dust density for dust disk density distributions.  The radial distributions commence (i.e. become non zero) at an \textit{inner truncation radius} specified by the parameter $R_{\rm tin}$, increase linearly to a peak at an \textit{inner radius}, $R_{\rm in}$, then decline exponentially to  truncate (become zero) at an \textit{outer truncation radius},  $R_{\rm t}$. The quantity $\chi$ parameterizes the slope of the portion of the distribution functions inside the inner radius. The parameter $h_0$, the \textit{radial scale length}, governs the rate of decline of the exponential disk portion of a distribution. The parameter $z_0$, the \textit{vertical scale height}, governs the rate of decline (exponential) of the distribution function measured from the model galaxy mid-plane. For stellar volume emissivity distributions it is important to note that, generally, the parameters $A_0$ and $h_0$ are wavelength dependent meaning the stellar emissivities themselves are wavelength dependent. 

The stellar volume emissivity distributions of stellar bulges in our model are described by S\'ersic profiles given by

\begin{equation}
{
    w(R,z) = w(0,0) \sqrt{ \frac{b_s}{2 \pi}} \frac{(a/b)}{R_{\rm eff}}
    \eta^{(1/2n_s)-1}
    \exp(-b_s \eta^{1/n_s}) 
}
\end{equation}

\noindent
with

\begin{equation}
    \eta (R, z) = \frac{ \sqrt{R^2 + z^2 (a/b)^2} } {R_{\rm eff}}.
\end{equation}

\noindent
Here ($a/b$) is the axial ratio, $n_{\rm s}$ the S\'ersic index, R$_{\rm eff}$ the effective bulge radius, and $b_{\rm s}$ a constant.

Examples of radial profiles of g-band stellar emissivity functions for disks and bulges are given in Fig.~\ref{fig:radial_distribution}. The profiles are for $z=0$. These examples correspond to the best fit model of NGC~3938, as described later in the paper.

\subsection{Model parameters}

Our model parameters are either geometric in nature, i.e. parameters controlling the variation of the analytic functions describing the intrinsic stellar luminosity density and dust density distributions, or parameters describing the amplitudes of such functions. The function amplitudes themselves, for a given geometry, are linearly proportional to the integrated total intrinsic (un-reddened) stellar luminosity density of a disk in the case of a stellar luminosity density distribution, or the peak dust optical depths in the case of dust density distribution. The dust optical depth parameterization we use is the B-band face-on galaxy optical depth measured at a dust disk inner radius, $\tau^f_B$. In turn, we note this optical depth, for a given dust disk geometry, is proportional to the total dust mass of its corresponding dust disk (see \citetalias{TP20}, Appendix B). Thus, in essence, we fit model SB profiles to the amplitude and shape of the data derived SB profiles by varying the integrated stellar luminosity density, the dust mass, and the geometric structure of each disk. From these fitted intrinsic quantities which fundamentally describe the state of a galaxy (sometimes coupled with additional assumptions), we can determine (in sections below) numerous fundamental measures used to understand the nature of our modeled galaxies.

We note the fact that our fitting process, constraining the intrinsic luminosity spectral density distribution of a galaxy disk or bulge, is a key advantage of our model over other models which rely on stellar population synthesis models to describe their stellar populations. By contrast, with our model the \textit{actual} intrinsic stellar luminosity function is derived \textit{from the data}. The latter fact, we point out, has potentially broader application for the characterization of galaxy stellar populations which we defer to future studies.

To reduce the number of free parameters and facilitate the fitting process, we set certain parameters to fixed values wherever possible. The fixed values are of two types: those determined from data and those constrained by the model on physically well-justified grounds. 

 \begin{table*}
\caption{Best fit values of geometrical parameters for the NGC 3938 and M101 RT models. All scale length and distance parameters have units of kpc.
}
  \begin{tabular}{ | ll | c }
    \hline
 \multicolumn{3}{c}{{\bf NGC 3938}} 
 \\ [-2ex]
 \multicolumn{3}{c}{}    \\
disk scalelength & $h_{\rm s}^{\rm disk}(ugrizJHKI_1I_2I_3$)  & (3.0, 3.0, 2.9, 2.9, 2.6, 2.5, .5, 2.5, 2.8, 2.8, 2.8)$\pm10$\% \\
thin disk scalelength & $h_{\rm s}^{\rm tdisk}$               &   $3.2\pm0.2$ \\
dust disk scalelength & $h_{\rm d}^{\rm disk}$                &   $9.0\pm^{0.7}_{0.3}$ \\
disk slope & $\chi_{\rm s}^{\rm disk}$                        &   $1.0\pm0.1$ \\
thin disk slope & $\chi_{\rm s}^{\rm tdisk}$                  &   $1.1\pm0.1$ \\
dust disk slope & $\chi_{\rm d}^{\rm disk}$                   &   $1.0\pm0.1$ \\
bulge effective radius & $R_{\rm eff}$                        &   $0.53\pm0.05$ \\
 
\hline\hline
\multicolumn{3}{c}{{\bf M101}} 
\\ [-2ex]
\multicolumn{3}{c}{}    \\
inner disk scalelength & $h_{\rm s}^{\rm i-disk} (u, g, r, i, J, K,I_1,I_2,I_3)$ & $(0.4, 0.4, 0.4, 0.4, 0.4, 0.4, 0.1, 0.1, 0.1)\pm 10\% $ \\
main disk scalelength & $h_{\rm s}^{\rm m-disk} (u, g, r, i, J, K,I_1,I_2,I_3)$ & $(3.7, 3.7, 3.7, 3.7, 3.3, 3.3, 4.0, 4.0, 4.2)\pm 10\% $\\
              \\
nuclear thin disk scalelength & $h_{\rm s}^{\rm n-tdisk}$ & $0.06$\\
inner thin disk scale-length &  $h_{\rm s}^{\rm i-tdisk}$ & $0.40\pm 0.04$\\
main thin disk scalelength &    $h_{\rm s}^{\rm m-tdisk}$ & $5.20\pm 0.52$\\
\\
inner dust disk scalelength &   $h_{\rm d}^{\rm i-disk}$ & $0.90\pm 0.18$\\
main dust disk scalelength &    $h_{\rm d}^{\rm m-disk}$ & $8.00\pm 1.60$\\
\\
inner and main disk slope &      $\chi_{\rm s}^{(\rm i-disk,m-disk)} $ & $(0.0, 0.1\pm 0.01)$\\
inner and main thin disk slope & $\chi_{\rm s}^{(\rm i-tdisk,m-tdisk)}$ & $(1.6\pm 0.16, 0.1\pm 0.01)$\\
inner and main dust disk slope & $\chi_{\rm d}^{(\rm i-disk,m-disk)}$ & $(0.0, 0.4\pm 0.04)$\\
bulge effective radius &         $R_{\rm eff}$  & $0.46\pm 0.05$\\
 \multicolumn{3}{c}{}    \\ [-1ex]
\hline
  \end{tabular}
  \label{tab:param_fitted} 
  \end{table*}

To illustrate the parameter reduction process we use the case of NGC 3938. The model for this galaxy consists of two morphological components: a central bulge and a main morphological component of disks. The main morphological component has the standard vertical stratification with the four concentric disks: the stellar disk, the thin stellar disk, the dust disk, and the thin dust disk.
The best fit geometric parameters are given in Table \ref{tab:param_fitted}, while the fixed geometric parameters are given in Table~\ref{tab:fixed_n3938}. 
The parameter reduction process consists of several parameter constraints. 
Thus, following \citetalias{PT11}, the scale length of the thin stellar disk was kept wavelength independent. From a physical point of view this assumption is justified from the fact that young stellar populations are physically associated with the molecular layer from where stars form. The thin stellar disk is formed by stars that have not had time to migrate from the thin molecular layer, and there is a strong spatial correlation with it. So there is no spread in age to allow for different slopes in the radial distributions as a function of wavelength. For the same reason, the thin dust disk, which is also physically associated with the molecular layer, is fixed to have the same spatial extent as the thin stellar disk. This means all the parameters of the thin dust disk, radial scale length $h_{\rm d}^{\rm tdisk}$, inner radius $R_{\rm in,d}^{\rm tdisk}$, inner truncation radius $R_{\rm tin,d}^{\rm tdisk}$, and inner slope $\chi_{\rm d}^{\rm tdisk}$ were fixed to those of the thin stellar disk. The thin stellar disk inner radii and inner truncation, $R_{\rm in,s}^{\rm tdisk}$ and $R_{\rm tin,s}^{\rm tdisk}$, were determined by visually inspecting the SB profiles of the data.

The vertical scale heights $z$ of all disks were fixed by the model as discussed in Sec. \ref{sec:model_descrp} above. 
GALFIT (Peng et al. \citeyear{2002AJ....124..266P}) fits to optical and 2MASS images determined that a S\'ersic index value of $n=2$ best represented the bulge stellar emissivity profile, value that also provided good fits for our model and was therefore adopted and fixed without further iterations.

More geometric parameters were required to model M101 than NGC 3938 due to its more complex structure. As will be discussed in \ref{results} below, three morphological components of disks and a bulge were found necessary to model this galaxy. The best fit geometric parameters are given in Table \ref{tab:param_fitted}, while the fixed geometric parameters are given in Table~\ref{tab:fixed_M101}.

The model fits to NGC 3938 and M101 were obtained using the optimisation procedure introduced in \citetalias{TP20} for modelling M33, and also  utilised to model NGC 628 (Rushton et al. \citeyear{2022MNRAS.514..113R}) and M51 (Inman et al. \citeyear{2023MNRAS.526..118I}).

\subsection{Fit quality evaluation} \label{sec:fitqual}

The quality of the SB fits were evaluated using the following metrics:

\begin{enumerate}

\item
The quantities  $D_{\lambda_l n}$[\%], which for given wavelength band of effective wavelength $\lambda_l$, are the absolute values of the percent deviations of each model predicted SB value from the observed SB value at the $n$\textquotesingle th annuli from the centre of a galaxy.

\begin{equation} \label{eq:R}
D_{\lambda_l n}[\%] = \frac{\overline{I_{\nu_l n}^{O}}-\overline{I_{\nu_l n}^{M}}}{\overline{I_{\nu_l n}^{O}}}\times 100
\end{equation} 
Here $\overline{I_{\nu_l n}^{O}}$ and $\overline{I_{\nu_l n}^{M}}$ are the azimuthally averaged observed and model SB values at the $n$\textquotesingle th discrete elliptical annulus, respectively, for the model $l$\textquotesingle th effective waveband of wavelength $\lambda_l$.
\\

\item The quantity $\overline{D_{\lambda_l}}$[\%], defined as the average value of \big|$D_{\lambda_l n}[\%]$\big| over all annuli of a SB profile at an effective waveband of wavelength $\lambda_l$
\begin{equation} \label{eq:Ravg} \overline{D_{\lambda_l}}[\%] = \frac{1}{N} \sum_{n=1}^N \big|D_{\lambda_l n}[\%]\big|.
\end{equation}
$ \overline{D_{\lambda_l}}[\%]$ measures the overall quality of the model predicted SB profile fit to the observed a SB profile at a particular waveband.
\\
\item Reduced chi$_{r \lambda_l}^2$ for a given waveband of effectve wavelength $\lambda_l$ given by
\begin{equation} \label{eq:chilam}
{\rm chi}_{\lambda_l}^2 = \frac{1}{N}\sum_{n=1}^N\frac{(\overline{I_{\nu_l n}^{O}}-\overline{I_{\nu_l n}^{M}})^2}{\epsilon_n^2}.
\end{equation}
In Eq. \ref{eq:chilam}, $\overline{I_{\nu_l n}^{O}}$ and $\overline{I_{\nu_l n}^{M}}$ are defined above, $\epsilon_n$ is the estimated surface brightness error in the $n$\textquotesingle th elliptical annulus (see Sec. \ref{sec:SB_prof}), and the sums are taken over all annuli with discernible galaxy emission to the background level at annulus N. Chi-squared measures the quality of the model predicted SB profile fit to the observed SB profile at a particular waveband of equivalent wavelength $\lambda_l$. Unlike the absolute deviation metrics discussed above, chi-squared accounts for noise in data, providing greater weight to data with less noise. Eq. \ref{eq:chilam} is approximate in that it assumes the number of data points in the defining sums are much greater than the number of model free parameters, a valid assumption in our case (see Inman et al. \citeyear{2023MNRAS.526..118I} for further discussion of this point).

\end{enumerate}

\begin{figure*}
    \includegraphics{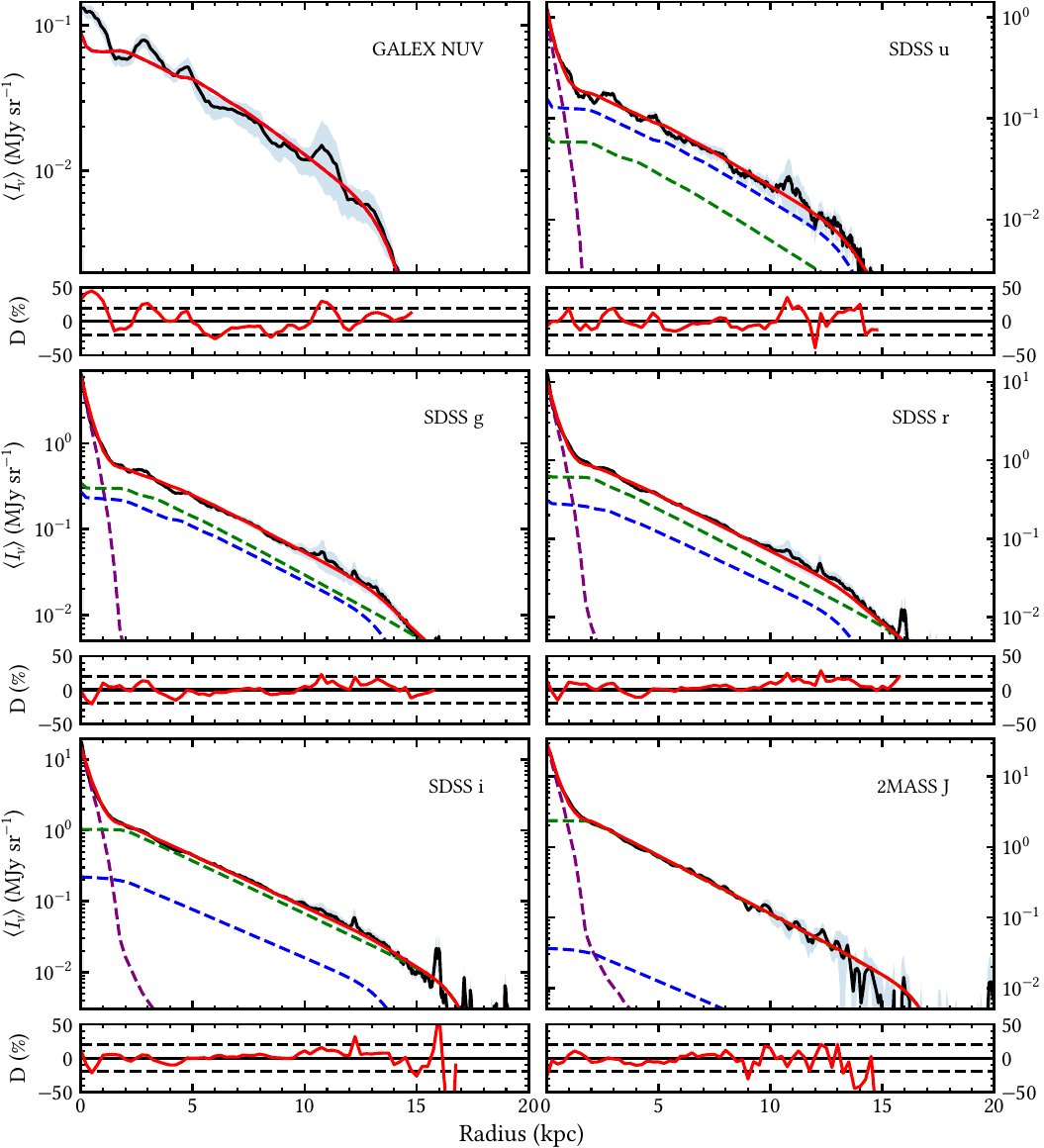}
    \caption{NGC 3938 azimuthally averaged SB profiles 
    in the UV, optical and NIR.
    The observed SB profile is plotted with the black curve with an error band indicated by blue shaded banding. The red curve plots the total SB of the best fit model profile. Surface brightness profiles from different model components contributing to the model total profile are also plotted with dashed lines: dark purple curve for the stellar bulge, blue curve for the thin stellar disk and green curve for the stellar disk. The percent differences between the best fit model SB values and the observed SB values, $D_{\lambda_l n}$[\%] (see Sec. $\ref{sec:fitqual}$), are plotted in the sub-plot below each profile, with the dotted black horizontal lines indicating +20\%, and -20\% deviation. All plotted SB values are corrected for MW foreground extinction as described in the text.
    }
    \label{fig:SB_UVOPT_N3938}
\end{figure*} 

\begin{figure*}
    \includegraphics{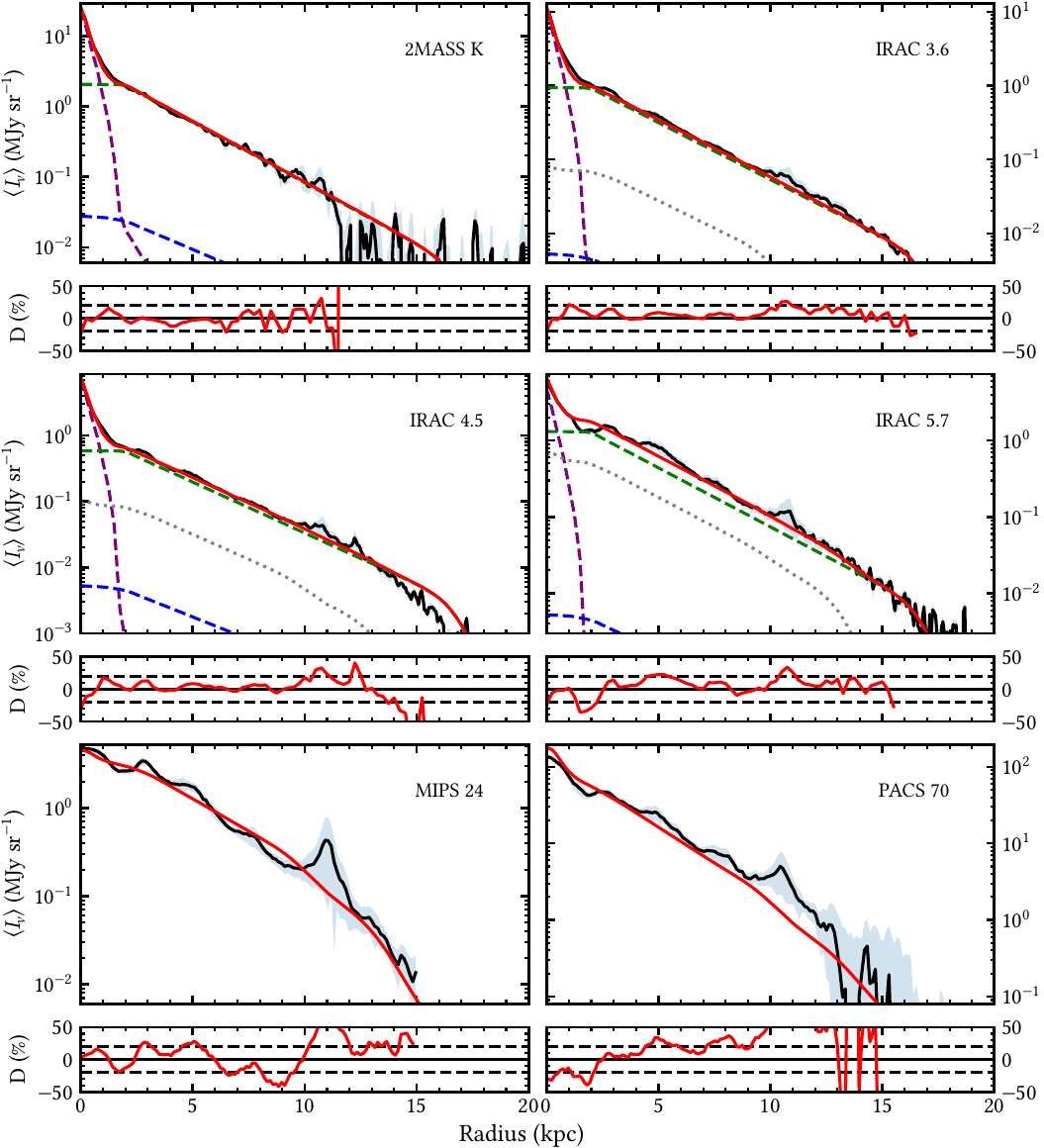}
    \caption{NGC 3938 azimuthally averaged SB profiles 
    in the NIR and MIR.
    The colour-coded plotted quantities and other plot details are the same as described in the caption to Fig. $\ref{fig:SB_UVOPT_N3938}$ except that the dust SB is also plotted with a grey-dotted curve.}
    \label{fig:SB_NIR_N3938}
\end{figure*}

\begin{figure*}
    \includegraphics{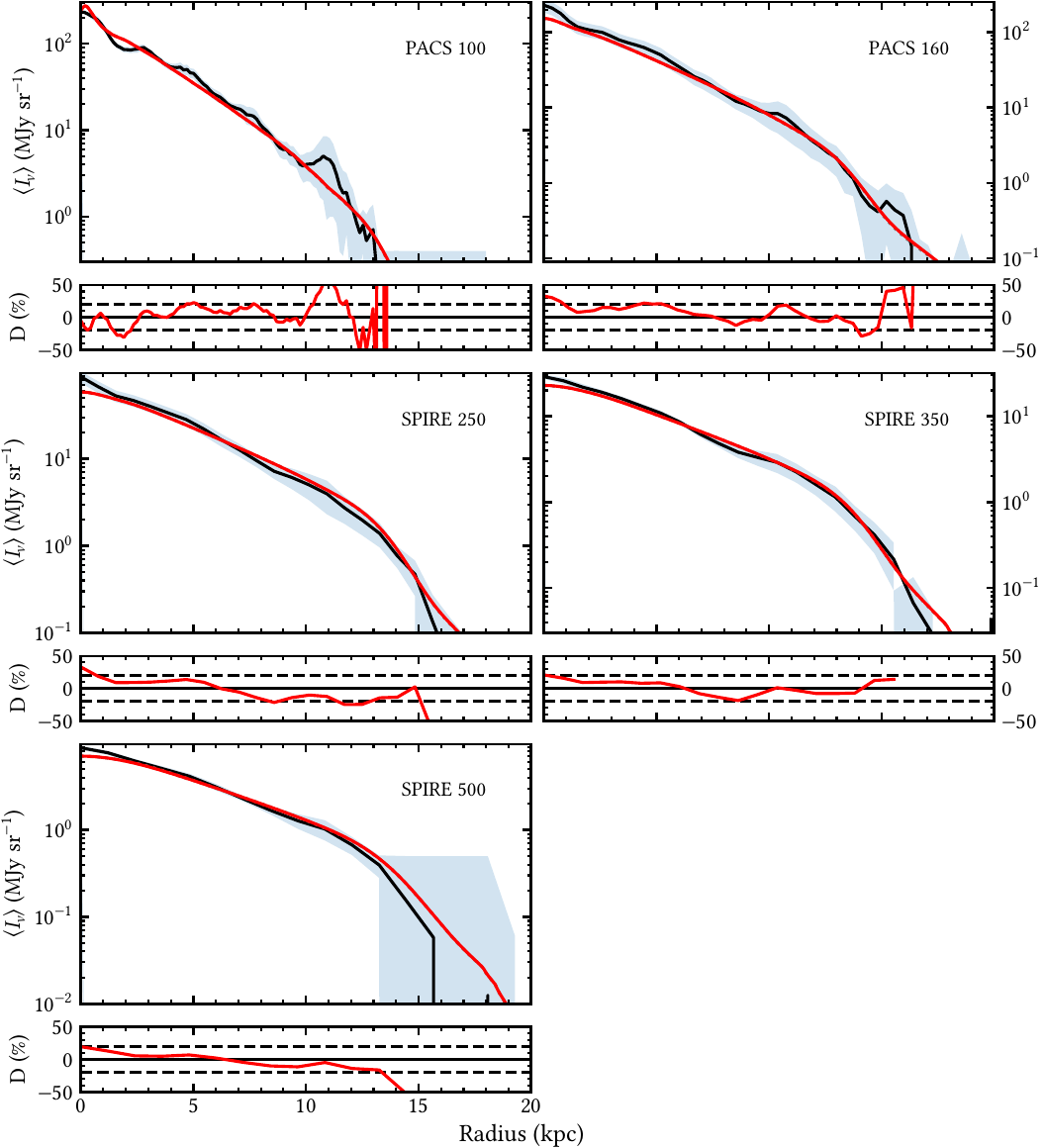}
    \caption{NGC 3938 azimuthally averaged SB profiles 
    in the FIR and submm. 
    The colour-coded plotted quantities and other plot details are the same as described in the caption to Fig. $\ref{fig:SB_NIR_N3938}$.}
    \label{fig:SB_IR_N3938}
\end{figure*}

\section{Results} 
\label{sec:Results}

\subsection{Azimuthally averaged surface brightness profiles} \label{results}

\subsubsection{NGC 3938} \label{sec:NGC 3938}

The SB profiles for NGC 3938 are displayed together with best fit model generated profiles in Figs. $\ref{fig:SB_UVOPT_N3938}$-$\ref{fig:SB_IR_N3938}$. We find the SB profiles are predominantly described by a single exponential function. However, in the optical/NIR, there are at least two breaks in the extended exponential profile. The first is a transition at $\approx 1.5$ kpc from the galaxy centre, which is due to the bulge and is seen on profiles from the 0.35 $\micron$ SDSS u-band to the 5.7 $\micron$ IRAC band. The second break is somewhat less pronounced on our profiles, but occurs at $\approx 13$ kpc. The latter break appears on our u, g, r, and i SDSS SB profiles, although this feature is absent 
in nature on profiles at longer wavelengths. The 13 kpc feature is well reproduced by our models, and is predicted to be due to the shorter truncation radius of the thin stellar disk with respect to the stellar disk, as discussed in Sect.~6.

Based on the overall characteristics of the SB profiles, we decided to model NGC 3938 with two morphological components, a bulge and a morphological component of disks. The required disks to fit the data were (as described in Sec. \ref{sec:model_descrp} above) a stellar disk, a thin stellar disk, a dust disk and a thin dust disk.

The geometric parameters for the best fitting model to the NGC 3938 SB profiles are given in Table $\ref{tab:param_fitted}$. Table \ref{tab:fit_quality} lists values for the fit quality metrics Eqs. \ref{eq:Ravg} and \ref{eq:chilam} at selected fitted wavebands. Our goal was to reduce the differences between best fit model and observed SB values at all wavelengths (Eq. \ref{eq:R}) to less than 20\%. As can be seen from Figs. \ref{fig:SB_UVOPT_N3938} - \ref{fig:SB_IR_N3938} this goal was largely met. The average value over all model wavelengths is 11.7\% consistent with our goal. 

It is instructive to understand in detail how emission from the different model disks contribute to the observed total SB profiles throughout the spectrum of wavelengths modeled. The GALEX NUV 0.22 $\micron$ SB profile panel shows the emission at this band is explained entirely by light from young population thin stellar disk stars (Fig. $\ref{fig:SB_UVOPT_N3938}$). The stellar disk, composed of a more evolved stellar population, begins to make a small emission contribution starting with the SDSS 0.36 $\micron$ u-band, although thin disk stellar emission remains dominant (Fig. $\ref{fig:SB_UVOPT_N3938}$). At the 0.48 $\micron$ SDSS g-band, the converse occurs, where emission from the stellar disk overtakes that of the thin stellar disk (Fig. $\ref{fig:SB_UVOPT_N3938}$). By the 0.77 $\micron$ SDSS i-band, stellar disk emission totally dominates the NGC 3938 surface brightness outside the bulge. At the 4.5 $\micron$ IRAC band, emission from the dust disks begin to make a perceptible contribution to the SB profiles, becoming increasingly important at 5.7 $\micron$ (Fig. $\ref{fig:SB_NIR_N3938}$). The NGC 3938 stellar bulge contribution to the total SB becomes apparent on the SDSS 0.36 $\micron$ u-band profile ($R_{\rm eff}$ = 0.53 kpc) evidencing statistically significant emission to the IRAC 4.5 and 5.7 $\micron$ bands (Fig. $\ref{fig:SB_NIR_N3938}$).

At wavebands longwards of 5.7 $\micron$, the SB emission is, for all practical purposes, completely from dust components (Fig. $\ref{fig:SB_IR_N3938}$). From 24 $\micron$ to 100 $\micron$ the observed SB profiles tend to be less smooth on small scales ($\approx 1$ kpc) than longer wavelength FIR SB profiles (Fig. $\ref{fig:SB_IR_N3938}$).  
These features are due to emission from localized HII regions with hot dust heated by young stars. 

The geometric best fit parameters of Table \ref{tab:param_fitted} and the fixed parameters from Table~\ref{tab:fixed_n3938} give further insight into the disk structure of NGC 3938. The model stellar volume emissivities and dust disk densities of NGC 3938 all truncate at the centre of the galaxy (parameters $R_{\rm tin, s}^{\rm disk}$,  $R_{\rm tin, d}^{\rm disk}$, $R_{\rm tin, s}^{\rm tdisk}$, $R_{\rm tin, d}^{\rm tdisk}$) and peak at a distance of 2 kpc from the centre (parameters $R_{\rm in, s}^{\rm disk}$, $R_{\rm in, d}^{\rm disk}$, $R_{\rm in, s}^{\rm tdisk}$, $R_{\rm in, d}^{\rm tdisk}$ ) - see Table~\ref{tab:fixed_n3938}. The radial scale lengths for the stellar disk volume emissivity functions decline gradually with wavelength from the largest value, $3.0\pm0.3$ kpc, at SDSS u-band to $2.5\pm0.2$ kpc at the K-band and then increases again to $2.8\pm0.2$ kpc in the IRAC bands (first row, Table $\ref{tab:param_fitted}$). The thin stellar disk scale length, $h^{\rm tdisk}_{\rm s} = 3.2\pm0.2$ kpc, is equal, within uncertainties, to the stellar disk u-band scale length. The trends in stellar disk scale lengths are consistent with the NGC 3938 SB profiles. For instance, on the SB profile panels of Figs. \ref{fig:SB_UVOPT_N3938} - \ref{fig:SB_NIR_N3938} stellar disk SB emission is seen to decline perceptibly faster than that of thin stellar disk emission. The dust disk density distribution scale length, $h^{\rm tdisk}_{\rm d} = 9.0$ kpc (Table $\ref{tab:param_fitted}$), is considerably larger than any stellar volume emissivity scale length. The truncation radii of the thin stellar and dust model disks are the same (14 kpc). The old stellar disc has a larger truncation radius of 17 kpc. 
We caution that truncation radii are particularly sensitivity dependent quantities, however, because in our model the truncation does not vary with wavelength, the truncations are usually derived at the wavelengths where they can be most reliably constrained. For example the 17 kpc truncation for the stellar disk is mainly fixed from the Spitzer 3.6 and 4.5\,$\mu$m. In the 2MASS bands the observed profiles seem to go into the noise at a shorter radius of around 11 kpc (see Fig.~\ref{fig:SB_NIR_N3938}). Nonetheless, with the fixed constraints of the model, the predicted profiles in the J and K band extend to 17 kpc - the model truncation of the stellar disk.

\begin{figure*}
    \includegraphics{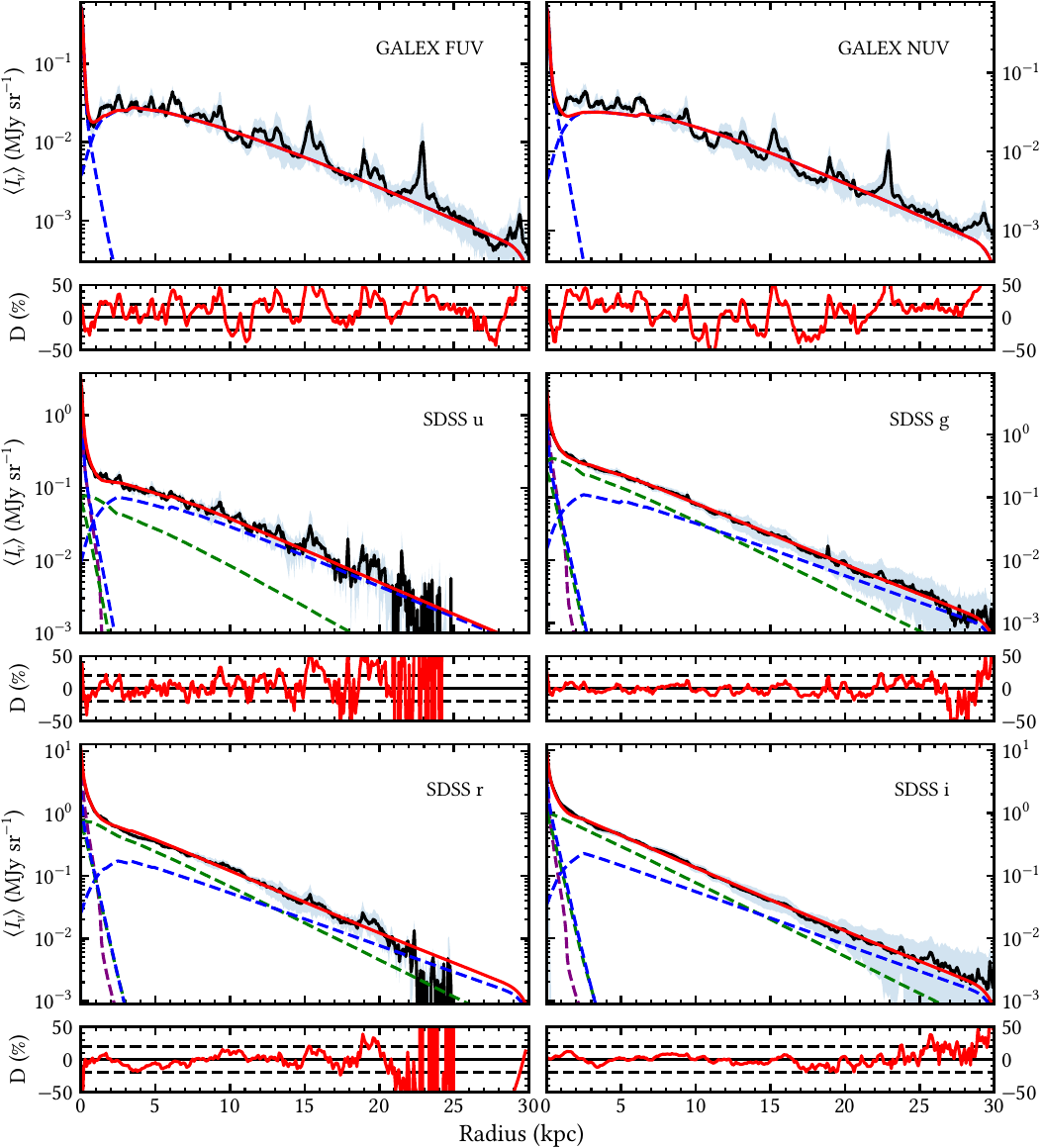}
    \caption{M101 azimuthally averaged SB profiles in the UV and optical. The observed SB profile is plotted with the black coloured curve with observational errors indicated by blue banding. The red curve plots the total SB of the best fit model profile. Surface brightness profile contributions to the total (red curve) from different model components are also plotted. The stellar bulge component SB contribution is plotted with the dark dashed purple curve. The thin stellar disk SB contributions (from the nuclear, inner and main thin disk) are plotted with a blue dashed curve. We note that we only show the sum of the nuclear and inner thin disks, to avoid overcrowding the figure. The stellar disk component SB contributions (from the inner and main disks) are plotted with a green dashed curve. The percent differences between the best fit model SB values from the observed SB values, $D_{\lambda_l n}$[\%] (see Sec. $\ref{sec:fitqual}$), are plotted in the sub-plot below each profile with the dotted black horizontal lines indicating +20\%, 0\%, and -20\% deviation. All plotted SB values are extinction corrected for MW foreground extinction as described in the text.
    }
    \label{fig:M101_SB_profile1}
\end{figure*}

\begin{figure*}
    \includegraphics{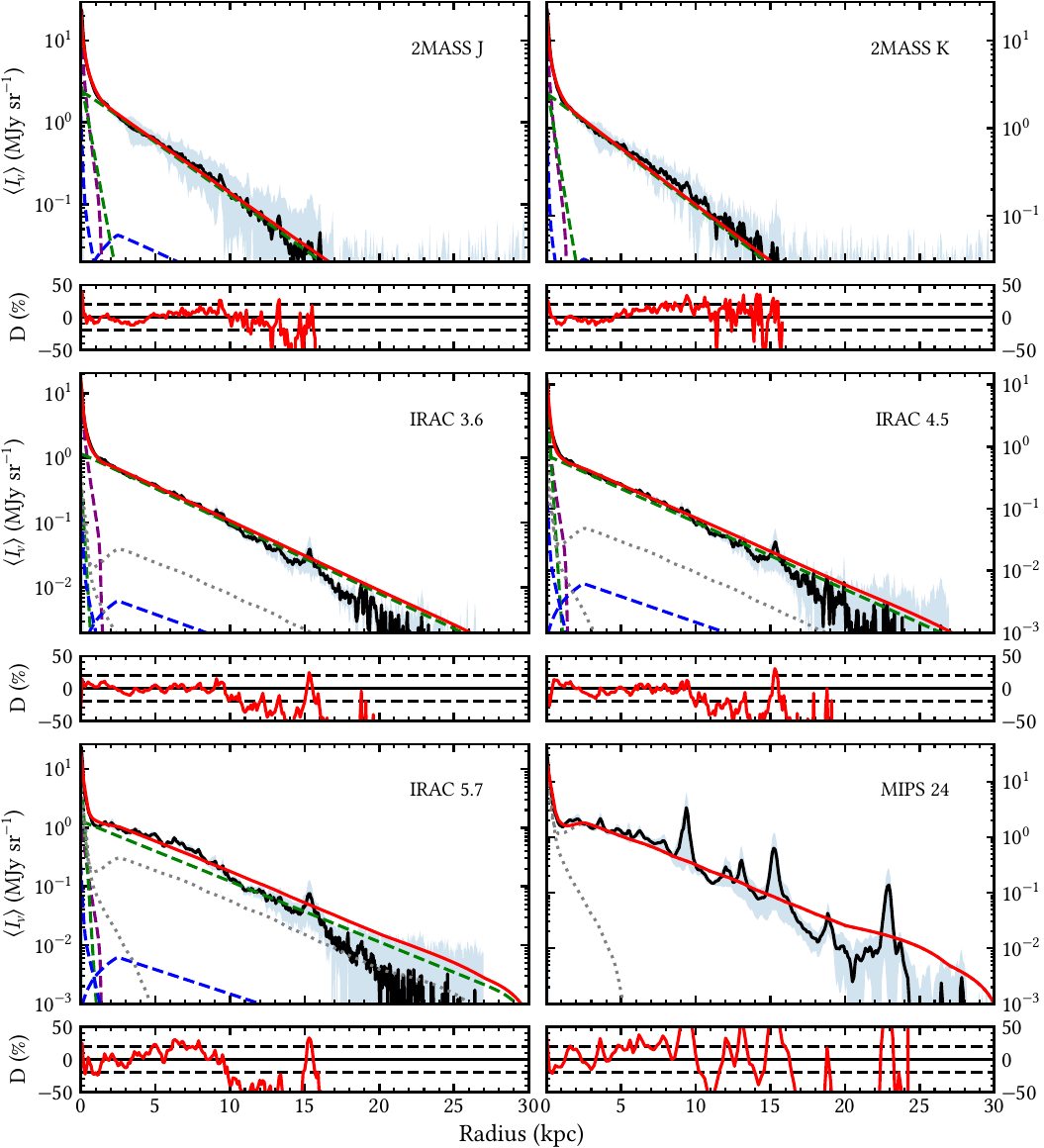}
    \caption{M101 azimuthally averaged SB profiles in the NIR and MIR. The colour coding of the profiles and other figure details are as described in the caption to Fig. \ref{fig:M101_SB_profile1}, except that the dust emission  is also plotted as a dotted grey line, making a non-negligible contribution to the total emission starting at 5.7 $\micron$.}
    \label{fig:M101_SB_profile2}
\end{figure*}

\begin{figure*}
    \includegraphics{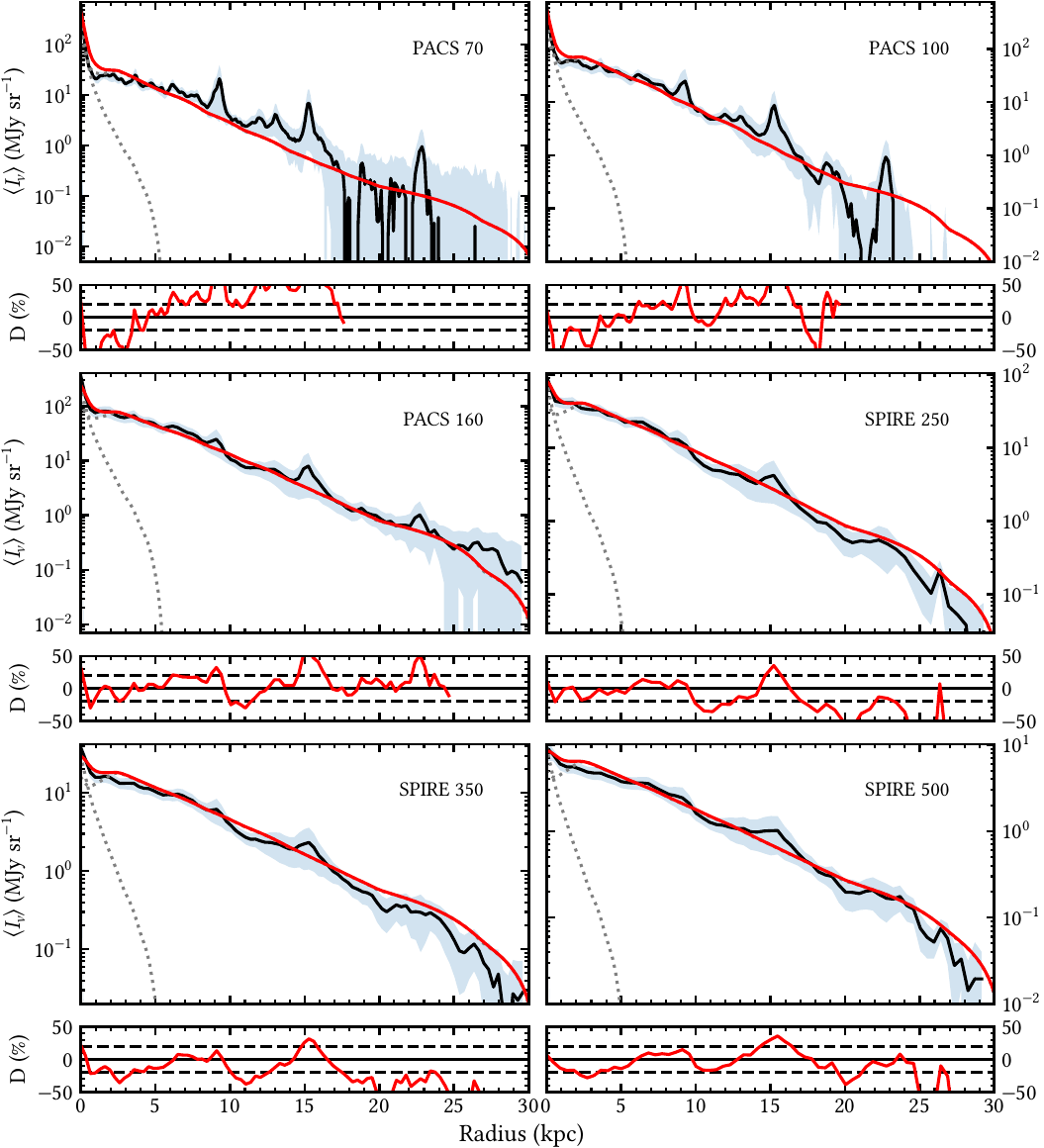}
    \caption{M101 azimuthally averaged SB profiles in the FIR and submm. 
    At these wavelengths the M101 emission is completely dust dominated. 
    The colour coding of the profiles and other figure details are as described in the caption to Fig. \ref{fig:M101_SB_profile2}.}
    \label{fig:M101_SB_profile3}
\end{figure*}

\subsubsection{M101}

The SB profiles for M101 are displayed together with best fit model generated profiles in Figs. $\ref{fig:M101_SB_profile1}$-$\ref{fig:M101_SB_profile3}$. Table \ref{tab:param_fitted} holds the geometric parameters for the best fitting model, while Table \ref{tab:fixed_M101} lists the fixed parameters. In contrast to NGC 3938, we found that the model fit to the M101 SB profiles was best obtained by employing more than one morphological component, as follows: a central bulge, a nuclear component composed of only a thin stellar nuclear disk of young stars, an inner component, and a main component. The inner and main components have each the vertical stratification consisting of a stellar disk, a thin stellar disk, a dust disk, and a thin dust disk.

Table {\ref{tab:fit_quality}} lists the average of the absolute value difference between model and fit values for SB profiles (Eq. \ref{eq:Ravg}) at selected fitted wavelengths. The average value over all model wavebands is 9.7\%, below our fit goal of 20\%. 
As with NGC 3938, the spike-like deviations in the SB profiles seen on UV and 24 - 70 $\micron$ profiles are non-axisymmetric emission features caused by stellar UV emission and hot dust emission typically associated with regions of  massive star formation. Prominent among these features is a particularly strong spike on the GALEX and 24-160 $\micron$ profiles at $R\sim$ 22 kpc. This feature is caused by emission from the giant HII region NGC 5471, which is several times more luminous than 30 Dor and possibly containing a hypernova remnant (Sun et al. \citeyear{2012ApJ...760...61S}). Such non-axial features are clearly less evident on the optical and NIR SB profiles where emission is dominated by the more smoothly distributed (spatially) older stellar populations, and also at the very longest wavelengths, where emission is primarily from cool dust.

The emission contribution from the thin stellar disk dominates, as expected, at the GALEX FUV and NUV bands (Fig. \ref{fig:M101_SB_profile1}). At the 0.36 $\micron$ SDSS u-band, the  SB is still dominated by emission from its thin stellar disk, with the stellar disk making a small but perceptible contribution (Fig. \ref{fig:M101_SB_profile1}).
From the 0.48 $\micron$ SDSS g-band to the 0.77 $\micron$ SDSS i-band significant contributions to the SB are  from both the stellar disk and thin stellar disk, but with the thin stellar disk emission contribution dominating in the outer regions of the galaxy. From the 1.26 $\micron$ 2MASS J-band to the 2.2 $\micron$ 2MASS K-band the component SB is dominated by emission from the stellar disk. From the 3.6 $\micron$ IRAC band to the 5.7 $\micron$ IRAC band, both dust and disk stellar emission make up the total SB emission.

A particularly notable feature of the dust emission, evident starting at the 3.6 $\micron$ IRAC band and at all longer wavelength SB profiles (although somewhat less conspicuous at 500 $\micron$), is a strong central peak of dust emission (Figs. \ref{fig:M101_SB_profile2} - \ref{fig:M101_SB_profile3}). Our model indicates this peak is mainly due to dust emission from the inner disk.

The best fit geometric parameters of Table \ref{tab:param_fitted} and fixed parameters of Table~\ref{tab:fixed_M101} reveal the structure of the model dust and stellar disks of M101. The inner and main stellar disks, as well as the nuclear and inner thin stellar disks have their inner radii at M101\textquotesingle s centre (thus identical with M101\textquotesingle s inner truncation radii). The main thin stellar disk and the dust disk have an inner truncation radius of 2.5 kpc. The nuclear thin stellar disk extends to an outer truncation of 1.0 kpc, while the inner disk and thin disk extend to 5 kpc. The outer truncation radii of the main stellar and dust disks were fixed to 30 kpc, although in the GALEX and SDSS data there is evidence for faint and non-axi-symmetric emission beyond 30 kpc.
The existence of this outer disk emission is consistent with the results from the imaging studies of Mihos et al. \citeyear{2013ApJ...762...82M}
and Merritt et al. \citeyear{2016ApJ...830...62M}.
The Merritt et al. study was based on very deep CCD g and r-band imaging intended to reveal the outer halo structure of spiral galaxies and extends to $R\approx48$ kpc (corrected to our assumed distance). However, our lower resolution dust emission data is not sensitive to emission beyond 30 kpc. Since our model needs coverage over the whole spectral range, in particular in the FIR and submm, we are unable to constrain any such emission. Because of this we truncated the model to 30 kpc for both the stellar and dust disk components.

The best fit scale lengths of the stellar volume emissivity and dust density functions for the M101 model disks (Table \ref{tab:param_fitted}) reveal useful information regarding the relative distribution of stellar populations and dust. The nuclear disk of young stars in M101 has a stellar volume emissivity scale length of only 60 pc and a truncation radius of 1 kpc, that is thus quite compact with respect to the overall size of M101. Interestingly the stellar volume emissivity scale lengths of all inner stellar disks at all wavelengths are the same, 0.4 kpc (first row, Table \ref{tab:param_fitted}). The inner dust disk density scale length, 0.9 kpc indicates a slower decline of dust disk density than thin stellar disk volume luminosity density with radius. The extended main stellar disk volume luminosity density, truncating at 30 kpc, is characterized by scale lengths decreasing with wavelength from 3.7 to 3.1 kpc (second row, Table \ref{tab:param_fitted}).  
The main thin stellar disk scale length is 5.2 kpc, indicating a slower decline of the volume luminosity density of the young stellar population with radius, as reflected on the SB profiles. 

\subsection{Volume stellar emissivity and volume dust density radial functions}

The best fit model parameters to the SB profiles discussed above determine the M101 and NGC 3938 volume stellar emissivity (W Hz$^{-1}$ pc$^{-3}$) and volume dust grain density (M$_{\odot}$ pc$^{-3}$)
for each sampled wavelength, at each $(R, z)$ point for each dust and stellar disk of the model. These fitted analytic functions, defined by model geometric fit parameters in Table \ref{tab:param_fitted} and fitted amplitude parameters, form the basis on which we calculate model intrinsic properties (e.g. disk luminosity densities, dust disk masses, dust disk optical depths, SEDs, etc.).
\begin{figure}
 \includegraphics{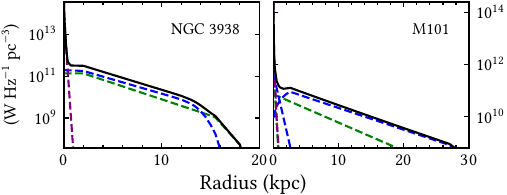}
    \caption{ Examples of radial profiles of g-band volume stellar emissivity at $z=0$ for the best-fit model for NGC 3938 (left) and M101 (right). Both panels display the total volume stellar emissivity (solid black line), as given by the sum of  the different model stellar components: bulge (purple line), stellar disk (green line) and thin stellar disk (blue line).}
    \label{fig:emissivity}
\end{figure}
\subsubsection{Volume stellar emissivity functions}

As an example, the volume stellar emissivities for NGC 3938 at the 0.48 $\micron$ SDSS g-band in the galactic mid-plane ($z=0$ kpc) for all stellar disks and the bulge are displayed in the left panel of Fig. \ref{fig:emissivity}.  
We can see that the main stellar disk volume emissivity is comparable to that of the main thin stellar disk and that both disks have similar scale lengths leading to similar declines rates, although the main thin stellar disk volume emissivity decline is slightly less. This behaviour is in accordance with the relative values of the respective scale lengths given of Table \ref{tab:param_fitted} (rows 1 and 2). In the inner parts of NGC 3938, the stellar emissivity from the bulge begins to make notable contribution shortwards of $R\approx$ 1.5 kpc. 
Our model predicts that the thin stellar disk and stellar disk truncate at $R=14$ kpc and $R=17$ kpc, respectively, suggesting that NGC 3938 may have a surrounding $\Delta$R $\approx$ 3 kpc annulus consisting mainly of evolved stars.

The right-hand panel of Fig. \ref{fig:emissivity} plots the mid-plane volume stellar emissivity functions for all stellar disks and the bulge for M101 at the 0.48 $\micron$ SDSS g-band to the outer truncation radius of 30 kpc. 
At $R\gtrapprox$ 2.5 kpc, the decline in the stellar g-band total emissivity (black curve) is exponential in character to the truncation radius.
We also see that our model predicts (at g-band in the mid-plane) the thin stellar disk to contribute significantly more to the total stellar emissivity of the main morphological component than does the stellar disk, completely dominating in the outer regions.  Notable as well is the longer scale length of the main thin stellar disk compared to that of the main stellar disk, numerically manifested by the difference in radial scale lengths (rows 2 and 5, Table \ref{tab:param_fitted}). Overall there is a contrasting difference in the behaviour of the stellar emissivity profiles between M101 and NGC 3938. 

Although the above discussion of the stellar emissivity functions is g-band specific and confined to the galactic planes, it nevertheless yields useful insight into the model stellar structures of NGC 3938 and M101 and provides a working and tangible example of how the model parameters of Table \ref{tab:param_fitted}
determine the intrinsic physical structure of these galaxies.

\subsubsection{Volume dust density functions}

\begin{figure}
 \includegraphics{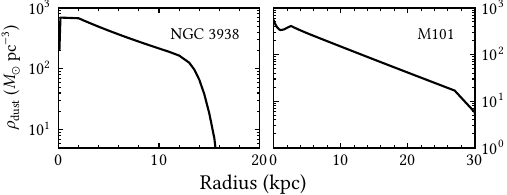}
    \caption{Dust density profiles for NGC 3938 (left) and M101 (right).}
    \label{fig:dust_density}
\end{figure}

The left-hand panel of Figure \ref{fig:dust_density} displays the radial profile of total dust density in the galactic plane ($z=0$ kpc) of NGC 3938. The dust density of NGC 3938 rises from the centre to the 
dust disk inner radius at 2 kpc ($R_{\rm in, d}^{\rm tdisk}$, Table \ref{tab:fixed_n3938}). From this point the dust density declines steadily to just short of the outer dust disk truncation radius at 14 kpc ($R_{\rm t,d}^{\rm disk}$).

The right-hand panel of Figure \ref{fig:dust_density} displays the radial profile of total dust density in the galactic plane ($z=0$ kpc) of M101. The profile shows a more complex structure than that of NGC 3938, reflecting  the different contributions of the inner and main morphological component dust disks. The dust density declines rapidly in the inner morphological component to a local minima at $R\approx1.5$ kpc, then rises to the inner radius of the main dust disk at $R=2.5$ kpc. After this radius, the dust density declines exponentially to the outer truncation radius of the main dust disk.

\begin{figure}
    \includegraphics[width=8.5cm]{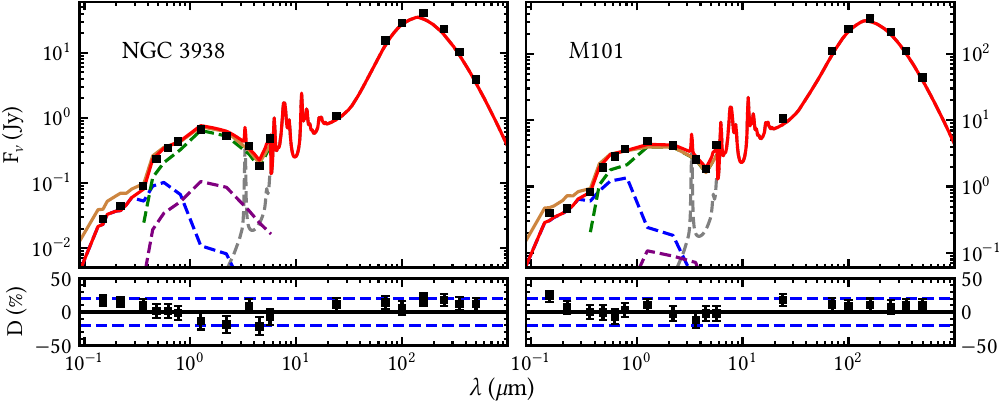}
    \caption{Spatially integrated model SED of NGC 3938 (left) and M101 (right) compared with observed data (black squares). Plotted on the horizontal axis is wavelength in $\micron$ and on the vertical axis flux density $F_{\nu}$ in Janskys. 
    The total model predicted flux density (sum of all constituent contributions) is plotted with solid red curves. The predicted flux density contribution from the different components is plotted as follows: dark purple for the bulge stars, light green for all stellar disk stars,  blue for all thin stellar disk stars, grey for all dust disks.
    The black square symbols on the lower panels of each SED plot the percent deviations of observed flux density values from best fit model predictions (i.e. from the solid black curves). To facilitate easy reference, the dotted horizontal lines on the lower panels of each plot mark percent deviation values of $\pm 20\%$.}
    \label{fig:SED_COMB}
    \end{figure}

\begin{figure}
    \includegraphics[width=8.5cm]{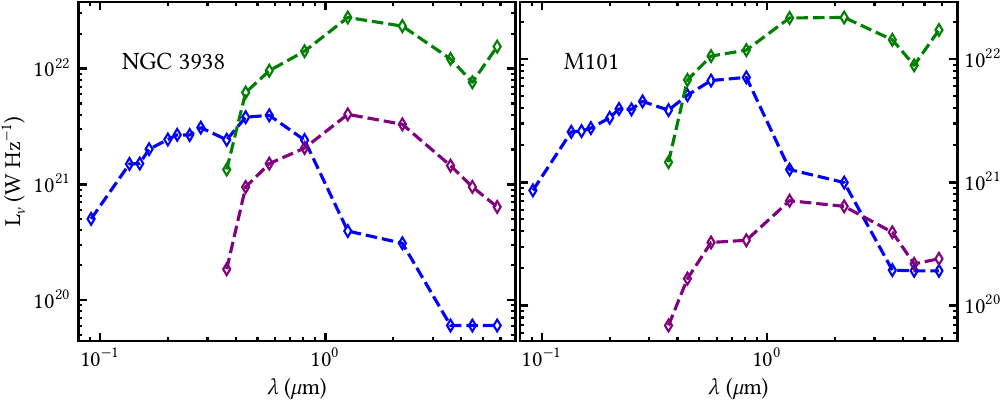}
    \caption{Intrinsic model SEDs for the stellar components. The colour code is as in Fig.~\ref{fig:SED_COMB}. The model predicted integrated stellar power density as a function of wavelength is plotted for each stellar component of NGC 3938 and M101.The plotted values for NGC 3938 are taken directly from the relevant columns of Table~\ref{tab:intr_lumin_N3938}. For M101, the plotted values are those tabulated for the bulge, the sum of all thin stellar disk luminosities densities at a particular $\lambda$, and the sum of stellar disk luminosity densities at a particular $\lambda$, with all values taken from Table~\ref{tab:intr_lumin_M101}.}
    \label{fig:intrinsic_SEDs}
\end{figure}

\begin{figure}
    \includegraphics{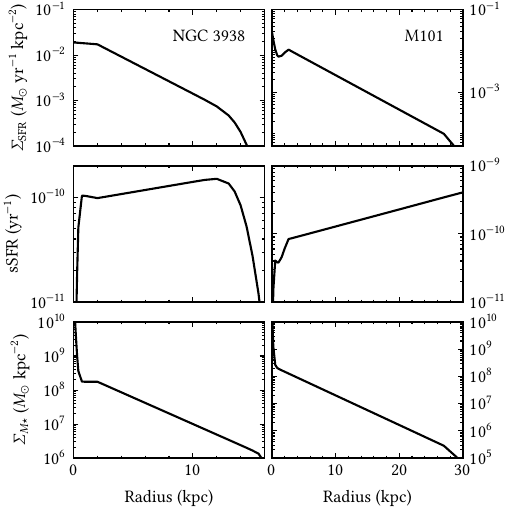}
    \caption{From top to bottom, radial profiles of SFR surface density ($\Sigma_{\rm SFR}$), specific star formation rate (sSFR), and stellar surface mass density ($\rm \Sigma_{M_{\star}}$) as a function of distance from the centre for NGC 3938 (left panel set) and M101 (right panel set) for the best fit model.}
    \label{fig:N3938_M101_eta}
\end{figure}

\subsection{Spectral energy distributions (SEDs}

The predicted observed spatially integrated SEDs of the best fit models of 
NGC 3938 and M101 are displayed in Fig.~\ref{fig:SED_COMB}. The SEDs are plotted with constituent disk contributions shown, along with observed flux density values. The percent residual deviations (D [\%]) of the observed flux density values from  model predictions are shown on the lower panel of each SED plot with dotted horizontal lines marking $\pm20\%$ levels to aid error interpretation. As can be seen, the predicted SEDs for both galaxies are in excellent agreement with observational SEDs. The average of the absolute values of residuals in the lower panels are 11.72\% for NGC~3938, and 9.75\% for M101. 

\begin{table}
    \caption{Intrinsic stellar luminosity and dust emission luminosity contributions in [W] for NGC 3938 and M101.  $L^{\rm total}_{s}$ is the total stellar luminosity, $L^{\rm tdisk}_{s}$ is the luminosity for all thin disk stars, $L^{\rm disk}_{s}$ is the luminosity of all stellar disk stars, $L^{\rm bulge}_{s}$ is the luminosity of all the bulge stars, and $L^{\rm tot}_{\rm d}$ is the total luminosity of the dust emission.
    }
    \begin{tabular}{lll}
        \hline
        & NGC 3938 & M101 \\
        \hline
        $L_{\rm s}^{\rm total}$ & $1.63\times 10^{37}$ & $1.95\times10^{37}$ \\
        $L_{\rm s}^{\rm tdisk}$ & $6.03\times 10^{36}$ & $9.75\times10^{36}$ \\
        $L_{\rm s}^{\rm disk}$ & $8.75\times 10^{36}$  & $9.12\times10^{36}$ \\
        $L_{\rm s}^{\rm bulge}$ & $1.49\times 10^{36}$ & $2.28\times10^{35}$ \\
        $L_{\rm d}^{\rm total}$ & $5.51\times 10^{36}$ & $6.33\times10^{36}$ \\
        \hline
    \end{tabular}
    \label{tab:lumin_stellar}
\end{table}

The intrinsic luminosities of the stellar and dust model components for both galaxies are given in Table~\ref{tab:lumin_stellar} (stellar) and the intrinsic luminosity densities for each stellar component at each sampled wavelength are given in Table~\ref{tab:intr_lumin_N3938} and Table~\ref{tab:intr_lumin_M101} for NGC 3938 and M101, respectively.

We can compare our model luminosity values in Table \ref{tab:lumin_stellar} with those of other works. A particularly useful reference for this is the DustPedia Web archive of local galaxy properties which lists, among other quantities, both bolometric luminosities and dust luminosities for 875 local galaxies, including M101 and NGC 3938 (Davies et al. \citeyear{2017PASP..129d4102D}). Dust luminosities were determined by fitting Modified Black Bodies (MBB) models with a THEMIS dust model (Jones et al. \citeyear{2013A&A...558A..62J}; K{\"o}hler et al. \citeyear{2014A&A...565L...9K}; Jones et al. \citeyear{2017A&A...602A..46J}), while bolometric and stellar luminosities were determined (Nersesian et al. \citeyear{2019A&A...624A..80N}) by fitting CIGALE energy balances codes (Boquien et al. \citeyear{2019A&A...622A.103B}; Clark et al. \citeyear{2018A&A...609A..37C}; Noll et al. \citeyear{2009A&A...507.1793N}) using both the THEMIS  and Draine et al (\citeyear{2007ApJ...663..866D}; \citeyear{2014ApJ...780..172D}) (hereafter DL14) dust models to observed galaxy SEDs spanning NUV to sub-millimeter wavelengths. For comparison with our results we have scaled all the  DustPedia luminosities discussed in this paper to our assumed distance to M101 and 
NGC~3938 and expressed their units in Watts.

For M101, the DustPedia bolometric luminosities are $2.06\pm0.10\times10^{37}$ W and $2.11\pm0.11\times10^{37}$ W, for the CIGALE/THEMIS and CIGALE/DL14 model fits, respectively. 
These values compare well with our determination of $1.95\pm 0.10\times10^{37}$ W for the M101 intrinsic total stellar luminosity ($L_{\rm s}^{\rm total}$, Table \ref{tab:lumin_stellar}).

For NGC 3938, the DustPedia bolometric luminosities are $2.00\pm0.10\times10^{37}$ W and $1.97\pm0.10\times10^{37}$ W, for the CIGALE/THEMIS and CIGALE/DL14 model fits.  For NGC 3938 our model gives intrinsic stellar luminosity of $1.63\pm0.10\times10^{37}$ W, slightly lower than the DustPedia values ($L_{\rm s}^{\rm total}$, Table \ref{tab:lumin_stellar}).

For M101, DustPedia dust luminosities are $5.28\times10^{36}$ W, $7.11\pm0.36\times10^{36}$, and $7.21\pm0.36\times10^{36}$ W for the MBB/THEMIS, CIGALE/THEMIS and CIGALE/DL14 model fit, respectively. Aniano et al. \citeyear{2020ApJ...889..150A} found a dust luminosity of $7.77\pm0.23\times10^{36}$ W for M101 in their analysis of Herschel and Spitzer KINGFISH Survey (Kennicutt et al. \citeyear{2011PASP..123.1347K}) galaxy images using their MIPS 160 PSF results. 
These values bracket our dust luminosity of $6.33\times10^{36}$ W for M101 (Table \ref{tab:lumin_stellar}), with the differences reflecting variations in the dust model used and in the goodness of the fit between different models.

For NGC 3938 the DustPedia archive gives dust luminosities of $4.52\times10^{36}$ W, $6.16\pm0.62\times10^{36}$ W, and $5.77\pm0.43\times10^{36}$ W for the MBB, CIGALE/THEMIS, and CIGALE/DL14 models, respectively. Aniano et al. \citeyear{2020ApJ...889..150A} measured a dust luminosity for NGC 3938 of $6.32\pm0.38\times10^{36}$W.
Our determination of the NGC 3938 dust luminosity of $5.51\times10^{36}$ W (Table \ref{tab:lumin_stellar}) is consistent with the DustPedia CIGALE/DL14 model.

Figure \ref{fig:intrinsic_SEDs} plots the intrinsic stellar SEDs for NGC 3938 and M101 as they would appear in the absence of dust. These plots yield insight into the contribution to the total luminous power output as a function of UV-optical-NIR wavelength from the differing stellar components of these galaxies. The values plotted are taken from Tables \ref{tab:intr_lumin_N3938} and \ref{tab:intr_lumin_M101}. For both NGC 3938 and M101, the stellar disk luminosity density (green dashed curve) significantly exceeds that of the bulge stars (purple dashed curve). The stellar disk and bulge colour trends are quite similar. 

The stellar disk emission for NGC 3938 significantly exceeds that of the thin stellar disk emission at all wavelengths until $\approx0.35$ $\micron$. By contrast, in M101 the luminosity density contribution from thin stellar disks is elevated in the optical compared to NGC~3938.

\subsection{Star formation}
\label{sec:sfr}

\subsubsection{Total star formation rates}

The total star formation rates (SFRs) were determined from the integrated intrinsic UV luminosities calculated using Eq.~17 of \citetalias{PT11}. We find values of ${\rm SFR=3.2\pm0.2}$ ${\rm M}_{\odot}$\,yr$^{-1}$ and ${\rm SFR=2.2\pm0.1}$ ${\rm M}_{\odot}$\,yr$^{-1}$ for M101 and NGC~3938, respectively (Table \ref{tab:physparms}).

Our SFR value for M101 of $3.2\pm0.2$ M$_{\odot}$\,yr$^{-1}$ is significantly higher than the value of 2.33 M$_{\odot}$\,yr$^{-1}$ from Kennicutt et al. \citeyear{2011PASP..123.1347K}, derived by combining 24 $\micron$ and H$\alpha$ luminosities. The SED fits of Nersesian et al. \citeyear{2019A&A...624A..80N} for M101 give SFR values (as listed in the Dustpedia database; http://dustpedia.astro.noa.gr/; Davies et al. \citeyear{2017PASP..129d4102D}) of $4.8\pm0.2$ and $4.9\pm0.3$ for their CIGALE/THEMIS and CIGALE/Draine et al. (\citeyear{2007ApJ...663..866D}; \citeyear{2014ApJ...780..172D}) energy balance/dust models, respectively. These values are much higher than our derived SFR. Jarret et al. \citeyear{2019ApJS..245...25J} find a SFR value of $3.7\pm0.9$  M$_{\odot}$\,yr$^{-1}$  (re-scaled to our assumed distance) based on their preferred total WISE 12 $\micron$ SFR calibration of Cluver et al. \citeyear{2017ApJ...850...68C}. 
Within uncertainties, our SFR is consistent with the value from Jarret et al. \citeyear{2019ApJS..245...25J}. It is interesting to see that, even for a well known nearby galaxy, the spread in literature values for SFR is up to a factor of 2. Perhaps this is not unexpected, since literature values are based on empirical calibrations or energy balance models, that usually work well on a statistical basis rather than giving precise values for individual objects. Although RT models are also prone to uncertainties in the underlying physics of the ISM, we believe that they provide a more accurate solution for the SFR.

For NGC 3938, our SFR value determination is $2.2\pm0.1$ M$_{\odot}$\,yr$^{-1}$.
 As in the case of M101 above, we may compare this with the determinations of Kennicutt et al. \citeyear{2011PASP..123.1347K}, and the Nersesian et al. \citeyear{2019A&A...624A..80N} values from both their CIGALE/THEMIS and CIGALE/DL14 fits giving SFR values of 1.77 M$_{\odot}$\,yr$^{-1}$, $4.1\pm0.5$ M$_{\odot}$\,yr$^{-1}$, and  $3.7\pm0.4$ M$_{\odot}$\,yr$^{-1}$, respectively. As before, our derived SFR is significantly lower than the values in Nersesian et al. \citeyear{2019A&A...624A..80N} and higher than the value derived in Kennicutt et al. \citeyear{2011PASP..123.1347K}.

\subsubsection{SFR surface density profiles} \label{sec:SFR_surf_dens}

Star formation rate surface density ($\Sigma_{\rm SFR}$) profiles can be calculated from our model and provide a measure of the rate of star formation in a galaxy as a function of radial distance from the centre.

The top left panel of Fig.~\ref{fig:N3938_M101_eta} shows the radial profile of $\Sigma_{\rm SFR}$ of NGC 3938. 
The profile declines approximately exponentially from the galaxy centre to the thin disk truncation radius, 14 kpc (Table \ref{tab:fixed_n3938}). This behaviour is consistent with that of the 0.22 $\micron$ GALEX NUV band SB profile of Fig. \ref{fig:SB_NIR_N3938}, a rough tracer of recent star formation.

The right panel of Fig.~\ref{fig:N3938_M101_eta} displays the $\Sigma_{\rm SFR}$ of M101. A sharp spike in the SFR is seen at the galaxy centre due to the presence of the thin nuclear disk. Beyond the inner radius of the main thin disk (2.5 kpc) $\Sigma_{\rm SFR}$ declines exponentially to the truncation radius of the outer thin stellar disk (30 kpc). This decline is almost 3 orders of magnitude.

\subsubsection{Specific star formation rates}

Specific star formation rates (sSFR, SFR per unit mass) are of interest as they can be considered an inverse measure of the time scale necessary for stellar mass assembly and thus can potentially shed light on time scales for the formation of galaxy stellar structures. Dividing our SFR values by their corresponding estimated stellar mass values (see Sec. \ref{sec:Stellar mass} below) gives average sSFR values of $5.8^{+2.5}_{-1.3}\times10^{-11}$ yr$^{-1}$ and $5.0^{+0.57}_{-0.31}\times10^{-11}$ yr$^{-1}$ for M101 and NGC 3938, respectively. These are tabulated in Table \ref{tab:physparms}.

Radial profiles of sSFR  for NGC~3938 and M101 were obtained from our model by dividing SFR surface mass density profiles by stellar surface mass density profiles (described in Sec. \ref{sec:Stellar mass} below). These are displayed for both galaxies in the middle panels of Fig. \ref{fig:N3938_M101_eta}. The NGC 3938 sSFR is more or less constant throughout its extent, with a slight increase towards larger radii. In the centre there is a sharp drop-off.

This sSFR profile behaviour is understandable in terms of the SFR surface density and stellar mass surface density profiles of Fig. \ref{fig:N3938_M101_eta}. For instance the sharp decrease in surface mass density outwards from the NGC 3938 centre to 1 kpc without a corresponding sharp decrease in SFR surface density, explains the sharp increase in sSFR from the centre to 1 kpc. Past 1 kpc, both the SFR surface density and stellar mass surface density decline exponentially to their respective outer truncation radii, although with slightly different slopes. This dictates an almost  constant sSFR with radius past 1 kpc. 

The model predicted M101 sSFR (Fig. \ref{fig:N3938_M101_eta}) shows a sharp drop at the very centre of M101.
Beyond $R=2.5$ kpc, the inner radius of the main thin stellar disk, the profile rises exponentially to the outer truncation radius ($R=30$ kpc).
Excluding the inner region of M101 with its rapidly varying and more complex sSFR profile behaviour, due to the combined effects of the nuclear thin disk and bulge, the sSFR rises almost a factor of 5 from the inner radius of the main disk ($R=2.5$ kpc) to the truncation radius of M101 ($R= 30$ kpc). This is in sharp contrast to the case of NGC 3938 where the increase from the inner disk radius to the point of outer disk truncation is a mild $\approx$ 30\%.

\begin{table*}
\caption{Key M101 and NGC 3938 physical parameters. The star formation rate (SFR), specific star formation rate ( $\overline{\textrm{sSFR}}$), total stellar mass (M$_\star$), dust mass ($M_{\rm d}$), and gas mass-to-dust mass ratio are listed for each galaxy (G/D).}
  \begin{tabular}{cccccc}
    \hline

Galaxy & SFR (M$_{\odot}$\,yr$^{-1}$) & $\overline{\textrm{sSFR}}$  & M$_\star$ (M$_{\odot}$) & $M_{\rm d}$ (M$_{\odot}$) & (G/D)\\
\hline
M101 & $3.2\pm0.2$ & $5.8^{+2.5}_{-1.3}\times$ 10$^{-11}$ yr$^{-1}$ & $5.5\pm0.52\times10^{10}$ & $1.04\pm0.05$ $\times$ $10^8$ & $135\pm12$  \vspace{.1 cm} \\

NGC 3938 & $2.2\pm0.1$    & $5.0^{+0.57}_{-0.31}\times$ 10$^{-11}$ yr$^{-1}$ & $4.3\pm0.37\times10^{10}$   & $5.15\pm0.20$ $\times$ $10^7$ & $148\pm12$ \\

\hline
 \end{tabular}
 \label{tab:physparms}
 \end{table*}

\subsection{Stellar mass}
\label{sec:Stellar mass}

Stellar masses are not direct outputs of our model. To derive them we use our model flux densities and external calibrations existing in the literature. Specifically, the total stellar masses of M101 and NGC 3938 were calculated using the M$_\star$/L calibration of Eskew et al. \citeyear{2012AJ....143..139E} and our model 3.6 $\micron$ and 4.5 $\micron$ flux densities. 
 These calculated stellar masses are $5.5\pm1.4\times10^{10}$ M$_{\odot}$ and $4.3\pm2.2\times10^{10}$ M$_{\odot}$ for M101 and NGC 3938, respectively, and are listed in Table \ref{tab:physparms} along with other key derived physical parameters. The uncertainty in our mass estimates is due to that of the Eskew et al. \citeyear{2012AJ....143..139E} calibration, $\approx \pm30\%$.
Stellar masses for M101 and NGC~3938 exist in the literature, and were derived as part of the Spitzer Survey of Stellar Structure in Galaxies (S$^{4}$G) 
(Sheth et al. \citeyear{2010PASP..122.1397S}, Munoz-Mateos et al. \citeyear{2013ApJ...771...59M}, Querejeta et al. \citeyear{2015ApJS..219....5Q})
and listed in their catalogue (\footnote{https://irsa.ipac.caltech.edu/data/SPITZER/S4G/}). Their values are $4.2 \times 10^{10}\,{\rm M}{\odot}$ and $3.4 \times 10^{10}\,{\rm M}{\odot}$ for M101 and NGC~3938, respectively. Since these values were derived using the same calibration as us, the small difference is explained by some difference in the flux density used for the 4.5\,${\mu}$m. We use a model prediction for the stellar emission, without any dust emission included, while the S$^4$G catalogue uses the total observed flux density. At 3.6\,${\mu}$m the flux densities are practically identical to those from the S$^4$G catalogue, as it should be, since the contribution from dust emission is essentially zero at this wavelength.

To study the spatial variation of stellar mass in M101 and NGC 3938, radial profiles of the stellar mass surface density ($\rm \Sigma_{M_{\star}}$) are displayed on the two lower panels of Fig. \ref{fig:N3938_M101_eta}. The NGC 3938 profile declines rapidly and non-exponentially from the centre due to bulge stars to $R\approx2$ kpc, the inner radius of the stellar disk, after which it declines exponentially to the outer truncation radius at $R=17$ kpc. The M101 $\rm \Sigma_{M_{\star}}$ profile is similar to NGC 3938 profile, rapidly and non-exponentially declining to $R\approx1$ kpc due to bulge stellar mass, after which it declines exponentially, terminating at outer truncation radius $R=30$ kpc.

\subsection{Dust mass and optical depth}
\label{sec:results_dust_mass}

The peak face-on B band dust optical depths, $\tau^f_B$ of model disks, occur at disk inner radii $R_{\rm in}$. These optical depths are calculated for all the dust disks of M101 and NGC 3938 by our RT model and are listed in Tables \ref{tab:tau_NGC 3938} and \ref{tab:tau_M101} for M101 and NGC 3938, respectively. The peak total $\tau^f_B$ of NGC 3938 is 1.64. The peak total $\tau^f_B$  for M101 is at the centre, 2.37, which is also the inner radius of the inner disk. A lesser $\tau^f_B$ secondary maximum of total optical depth of 1.85    is predicted for M101 at the overlapping inner radius of the main dust disk  (at 2.5 kpc).

The mass of all dust in each galaxy can be calculated by summing the dust mass for each disk found by substituting their individual $\tau^f_B$\textquotesingle s into equations B1-B6 of \citetalias{TP20}. Doing so, we obtain total dust masses of $M_{\rm d}$ of $1.04\pm0.05\times10^8$ M$_{\odot}$ and $5.15\pm0.20\times10^7$ M$_{\odot}$ for M101 and NGC 3938, respectively (Table \ref{tab:physparms}).

Our model dust mass for M101, $1.04\pm0.05$ $\times10^8$ M$_{\odot}$ (Table \ref{tab:physparms}), can be compared to the value $6.9\pm1.3\times10^7$ M$_{\odot}$ (M101) from Aniano et al. \citeyear{2020ApJ...889..150A} (their Table 9, MIPS160 PSF value) who employed the dust model of Draine \& Li \citeyear{2007ApJ...657..810D}. The study of Nersesian et al. \citeyear{2019A&A...624A..80N} found a dust mass of $1.0970\pm0.1078$ $\times$ $10^8$ M$_{\odot}$ by fitting the CIGALE energy balance model. Both these determinations used the dust grain models of Draine et al. \citeyear{2007ApJ...663..866D} and \citeyear{2014ApJ...780..172D}.

For NGC 3938, our model gives a dust mass of $5.15\pm0.2$ $\times$ $10^7$ M$_{\odot}$ to be compared with that of $5.2\pm1.4$ $\times$ 10$^7$ M$_{\odot}$ from Aniano et al. \citeyear{2020ApJ...889..150A} determined as above for M101. The Nersesian et al. \citeyear{2019A&A...624A..80N} study gave a dust mass for NGC 3938 of $7.6760\pm0.7566$ $\times$ $10^7$ M$_{\odot}$.

We may conclude that for both M101 and NGC 3938, our dust mass determinations are consistent with the determinations of both Aniano et al. \citeyear{2020ApJ...889..150A} and Nersesian et al. \citeyear{2019A&A...624A..80N}. 

\begin{table}
\caption{Face-on B-band dust optical depths for NGC 3938 at the inner radius.
}
\begin{tabular}{ l l  l l}
    \hline
    & $\tau^f_B$ & $R_{\rm in}$ (kpc)     \\ 
    \hline
    \hline
    total & 1.64 & 2.0 \\
    \hline
\end{tabular}
\label{tab:tau_NGC 3938}
\end{table}

\begin{table}
\caption{Face-on B-band dust optical depths for M101 at the inner radii of the inner and main morphological disk components.}
\begin{tabular}{ l l  l l}
    \hline
    & $\tau^f_B$    & $R_{\rm in}$ (kpc) & \\
    \hline
    \hline
    total & 2.37 & 0.0\\
    \hline
    \hline
    total & 1.85 & 2.5 \\
    \hline
    
    \hline
\end{tabular}
\label{tab:tau_M101}
\end{table}
\subsection{Gas-to-dust mass ratios}
\label{sec:gas_to_dust}

We can estimate the gas mass-to-dust mass ratio (G/D) for M101 using the H and H$_2$ gas masses from Aniano et al. \citeyear{2020ApJ...889..150A} including the contribution of heavy elements (assuming solar abundances) and our dust mass (Sec. \ref{sec:results_dust_mass}), to obtain (G/D) = $135\pm12$ (Table \ref{tab:physparms}). Repeating this calculation in exactly the same way for NGC 3938 we arrive at (G/D) = $148\pm12$ (Table \ref{tab:physparms}). These (G/D) values are both consistent with expectations for late-type spiral galaxies. For example we may compare this to values of (G/D) of 100 for the MW (Bohlin et al. \citeyear{1978ApJ...224..132B}) and 370  determined by Lianou et al. \citeyear{2019A&A...631A..38L} for their modeled sample of 413 local late-type and irregular galaxies. We note that the Lianou et al. \citeyear{2019A&A...631A..38L} (G/D) sample shows considerable scatter encompassing our measured values as well as that of the MW (see for example their Figure 8). 
\subsection{Attenuation curves}
\label{sec:atten_disc}

\begin{figure}
	\includegraphics[width=\columnwidth]{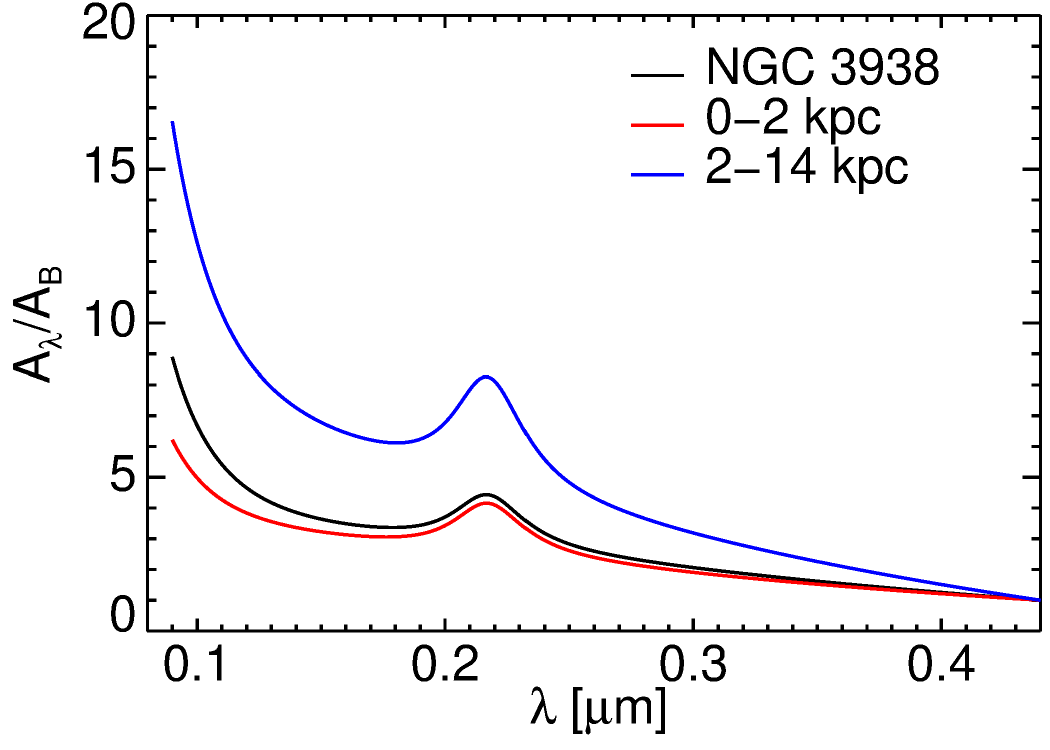}
    \caption{NGC 3938 attenuation curve. The global attenuation curve is plotted with the black solid line. Also plotted are the attenuation curves restricted to the inner (0-2 kpc) and outer (2-14 kpc) regions, with red and blue lines, respectively.}
    \label{fig:attencurve_N3938}
\end{figure}

\begin{figure}
	\includegraphics[width=\columnwidth]{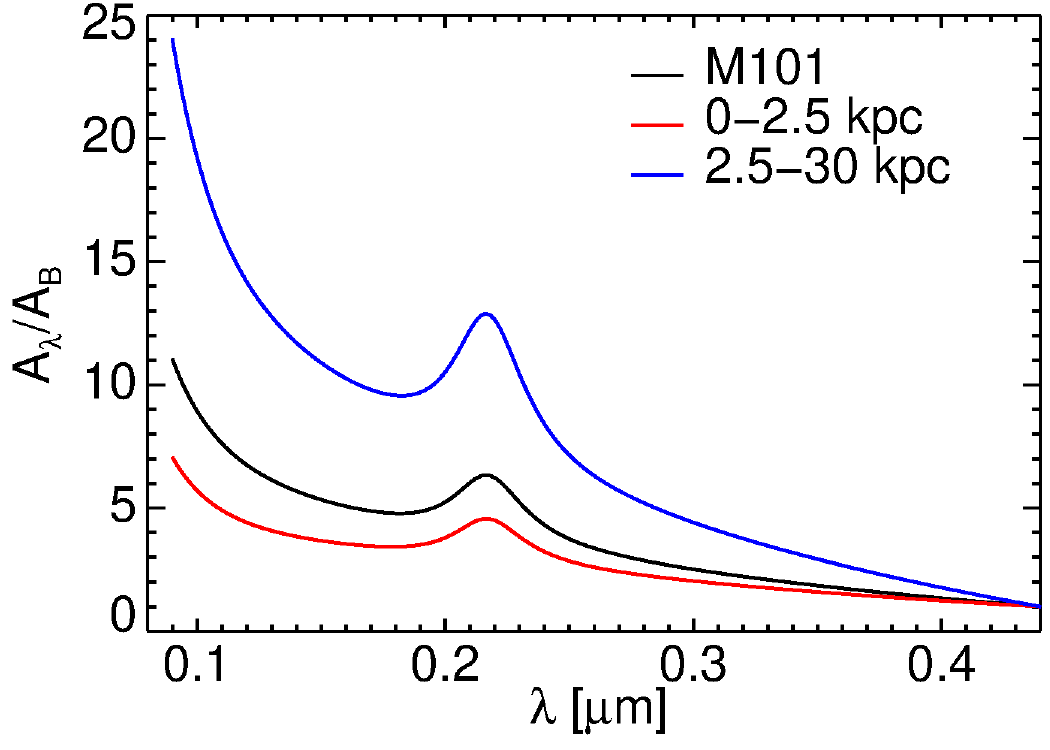}
    \caption{M101 attenuation curve.The global attenuation curve is plotted with the black solid line. Also plotted are the attenuation curves restricted to the inner (0-2.5 kpc) and outer (2.5-30 kpc) regions, with red and blue lines, respectively.}
    \label{fig:attencurve_M101}
\end{figure}

Our model facilitates the calculation of dust attenuation curves for stellar emission. From our model, we calculated attenuation values at all modelled bands.
Calculated attenuation values were then fitted to a parameterized attenuation function of Salim et al. \citeyear{2018ApJ...859...11S} (their Equations 8 and 9) to provide smooth attenuation curves. To gain insight into the dependency of dust attenuation on location within our modeled galaxies, we also calculated attenuation curves for the inner and outer portions of each galaxy separately. Curves were calculated for galaxy volumes interior to the inner radii of the main dust disk and for galaxy volume exterior to this point.  
The resulting attenuation curves are displayed in Figs.~\ref{fig:attencurve_N3938} and \ref{fig:attencurve_M101}. 
For both galaxies, the attenuation curves in the outer regions are significantly steeper than those in the inner regions. 

The attenuation curves for the entirety of NGC 3938 and M101 fall between those of the inner and outer galaxy values, although in the case of NGC 3938 the attenuation curve for the entire galaxy is closer to that of the inner region than for M101. 
The 2175 {\AA} UV "bump" is also clearly present on both galaxy\textquotesingle s attenuation curves and stronger in the outer galaxy for both galaxies.

The behaviour of the model attenuation curves of both galaxies are similar to those found for the spirals M33, NGC 628, and the MW found by applying the same RT model of this study (\citetalias{TP20}; Rushton et al. \citeyear{2022MNRAS.514..113R}; Natale et al. \citeyear{2022MNRAS.509.2339N}). Model results for these galaxies indicate attenuation curves increase more steeply at short wavelengths than long wavelengths when calculated for outer galaxy volumes as opposed to the inner volumes. For M33 \citetalias{TP20} attributed this behaviour to the relative optical depth difference between the inner and outer regions. This trend is also evident for both NGC 3938 and M101, and in fact more pronounced for M101 than NGC 3938, likely explainable by the paucity of dust in the inner $\approx$ 2 kpc of NGC 3938 compared to M101 (Fig. \ref{fig:dust_density}). The comparative lack of central dust in NGC 3938 with respect to M101 (Fig. \ref{fig:dust_density}) is also consistent with the fact that the total attenuation curve of NGC 3938 is closer to that of its inner attenuation curve than the same for M101 (Fig. \ref{fig:attencurve_N3938}). For both NGC 3938 and M101, the  2175 {\AA} UV bump increases in amplitude for attenuation curves made for stars in outer galaxy volumes as opposed to the inner galaxy volumes. A similar result is evident in the model attenuation curves of both the MW (Natale et al. \citeyear{2022MNRAS.509.2339N}) and M33 (\citetalias{TP20}).

\section{Discussion}
\label{sec:discussion}

In this section we will discuss and compare the model results obtained in Sec. \ref{sec:Results} above for NGC~3938 and M101 with each other. 
We will then put together the results for these two galaxies with our previous determinations for M33, NGC~628, M51 and the MW,  all obtained using the same RT models and procedures. We will look at the intrinsic properties derived for this small statistical sample of six galaxies.

\subsection{Intrinsic structural properties of NGC~3938 and M101}

Some aspects of the external structure of NGC 3938 and M101 are quite similar. For instance, both are late-type Sc spirals, M101 having distinctly traceable spiral arms and thus of the grand design type. NGC 3938 exhibits traceable spiral arms, but with a somewhat more complex structure and has been categorized as a multi-armed spiral by Korchagin et al. \citeyear{2005astro.ph..9708K}. Both galaxies also have bulges, but lack central bars. Nonetheless, there are also fundamental differences in their structure, as implied by our model fits.

In NGC~3938 we found that the stellar disk extends further  than the thin stellar disk and the dust disk. This leaves NGC~3938 with a shell of older stars in the outer regions. M101, on the other hand, is almost twice as extended and has the same outer truncation radius for all stellar and dust components of the main disk. 

Apart from their differences in extent, NGC 3938 and M101 also differ in complexity. NGC 3938 is described adequately by two morphological components: a central bulge component and one morphological disk component. By contrast, M101, with a much larger spatial extent, contains a nuclear disk, a bulge,  an inner disk and a main disk.

\subsection{SB photometry for NGC~3938 and M101: beyond standard B/D decomposition}

Our analysis allows for the decomposition of more than the standard bulge and disk morphological components. As mentioned before, we found the need to introduce a nuclear disk and inner disk for modelling M101, in addition to the main disk and the inner bulge. Furthermore, our model also allows for a vertical stratification, with stellar disks and thin stellar disks. Dust disks also have their own photometric parameters, that do not necessary coincide with the stellar parameters. Because of this our results are not directly  comparable with those from the literature. We therefore discuss in this section the closest association with existing literature values.

Our modelling of NGC 3938\textquotesingle s structure as a bulge with related exponential disks is consistent, from the point of view of number of morphological components,  with the early study of van der Kruit \& Shostak \citeyear{1982A&A...105..351V}. These authors performed photographic surface photometry of NGC 3938 at the photographic F-band (equivalent wavelength $\approx 6400$ \r{A}), nearby in wavelength to our analysed SDSS r-band data at 6200 \r{A}. These authors found an F-band scale length of 3.1 kpc (corrected to our assumed distance), consistent with our model r-band intrinsic stellar disk scale length of $h_{\rm s}^{\rm disk}=2.9\pm0.3$ kpc (row 1, Table \ref{tab:param_fitted}). This is in reasonable agreement particularly considering that these two quantities, while related, are not exactly the same. In our model $h_{\rm s}^{\rm disk}$ is the scale length of the stellar volume luminosity function for the stellar disk in the r-band past the inner radius, describing an intrinsic quantity, rather than the measured SB profile scale length of van der Kruit \& Shostak  \citeyear{1982A&A...105..351V}.  

Despite the dominant exponential nature of the SB profile found  for NGC 3938, two breaks are clearly evident in the  van der Kruit \& Shostak \citeyear{1982A&A...105..351V} F-band SB profile. The first break, is at $R\approx1.5$\,kpc. As noted above in Sec. \ref{sec:NGC 3938}, we see this break in our SB profiles from the u-band to 5.7 $\micron$, and it is the radial point where emission from the stellar disk begins to dominate that from the bulge. A second break in the F-band SB profile was noted by van der Kruit \& Shostak \citeyear{1982A&A...105..351V} at $R\approx13$\,kpc (assuming our distance). The latter break, is also present on our g, r, i, SDSS SB profiles (Fig. \ref{fig:SB_UVOPT_N3938}). 
van der Kruit \& Shostak \citeyear{1982A&A...105..351V} attributed this feature as being due to the decline of emission from old stellar disks, also seen in studies of edge-on galaxies by van der Kruit \& Searle (\citeyear{1981A&A....95..116V}, \citeyear{1981A&A....95..105V}) , which they refer to as "edges". In our model, NGC 3938 SB at optical wavelengths is due to a combination of emission from both a stellar disk truncated at 17 kpc and and a thin stellar disk truncated at 14 kpc, both having different scale-lengths. This produces the observed break at 13 kpc for all the bands where there is significant contributions from both disks. At longer optical/NIR wavelengths, where our model emission is dominated by the stellar disk only, we predict no such break, as reflected by the observations.

We should also mention the work of {{Castro-Rodr{\'\i}guez} \& {Garz{\'o}n}} \citeyear{2003A&A...411...55C}, who performed J and K$_s$-band imaging of NGC~3938 as part of an observational program to characterize the structural and photometric parameters of spiral galaxies. These authors fitted exponential profiles and a S\'ersic function to describe the bulge, finding exponential scale lengths of $2.577\pm0.0174$ kpc and $2.764\pm0.0260$ kpc at J and K$_s$, respectively, and half-light bulge radii of $0.547\pm0.026$ kpc and $0.593\pm0.017$ kpc at J and K, respectively, with our assumed distance. In addition,  Tully et al. \citeyear{1996AJ....112.2471T} imaged NGC 3938 as part of their study of the Ursa Major cluster, and determined exponential scale lengths 3.83 kpc, 3.44 kpc, 3.17 kpc, 2.70 kpc from their images made at B, R, I, and K, respectively assuming our distance. The above B, R, I, J and K band scale lengths may be compared with our intrinsic model stellar disk scale lengths of 
$3.0\pm0.3$ kpc, 
$2.9\pm0.3$ kpc, $2.9\pm0.3$ kpc, $2.6\pm0.2$ kpc, and $2.5\pm0.2$ kpc, at g, r, i, J, and K band, respectively (Table \ref{tab:param_fitted}).
At shorter wavelengths there is quite a strong disagreement. This could be due, in part, to the fact that previous studies did not take into account the effect of dust attenuation. Another reason may be due to the fact that, as discussed above, we use a more complex model, with both a stellar disk and a thin stellar disk, which only becomes relevant at shorter wavelengths. We note of course the improved agreement at longer wavelengths.
On the same note, the effective NGC 3938 bulge radii {{Castro-Rodr{\'\i}guez} \& {Garz{\'o}n}} \citeyear{2003A&A...411...55C} determined at J and K$_s$ above are quite consistent with our intrinsic model value of $0.53\pm0.05$ kpc (Table \ref{tab:param_fitted}).

For M101 SB photometry also exists in the literature. As part of their survey of halos around spiral galaxies, Merritt et al. \citeyear{2016ApJ...830...62M} imaged M101 in the SDSS filter g and r bands performing SB analysis. These authors fitted a bulge plus single exponential disk model to their g-band SB profile, obtaining a scale length for the disk of $3.9\pm0.1$ kpc (scaled to our assumed M101 distance). This observational value is 
difficult to compare with our determinations, since our model of M101 is by far more complex, containing more morphological components. Nonetheless, the literature value lies between our determination of the main stellar disk scale length at g of $3.7\pm0.3$ kpc, and of the main thin stellar disk scale length of $5.2\pm0.5$ kpc. 
Merritt et al. \citeyear{2016ApJ...830...62M} also determined an effective bulge radius of $1.5\pm0.3$ with a S\'ersic index $n = 1.9\substack{+0.5 \\ -0.4}$ to be compared with our model value of $0.46\pm0.05$ kpc with a S\'ersic index $n=2$. 
The reason for this substantial discrepancy could be related to comparative instrumental resolution issues. Merritt et al. binned their g-band images to 16.8 arcsec wide pixels before their SB analysis, which at our assumed distance amounts to 0.55 kpc on the sky. This choice provides spatial sampling slightly below the Nyquist limit generally considered necessary to resolve a 1.55 kpc feature. By contrast, typical g-band SDSS images (as used in our analysis) have typical seeing FWHM seeing profiles of $\approx 1.4$ arcsec, or 0.046 kpc on the sky, providing ample resolution to accurately sample the M101 bulge.
 However, despite the resolution issues, we believe that the main reason for this  discrepancy is again related to the fact that we used a more complex model, with a nuclear disk and an inner disk in addition to the bulge. It is likely that the bulge determination of Merritt et al. is affected by contribution from what we believe to be nuclear and inner disk. In other words, in the absence of a nuclear disk in the model, the central peak can only be fitted with a higher S\'ersic index for the bulge, and then the S\'ersic index and effective radius are correlated parameters. We note that our model analysis benefits by the additional constraints of the panchromatic analysis, including data from IRAC and GALEX, which allows a more detailed
decomposition of SB contributions from different morphological components. 

A more recent B/D decomposition for M101 has been performed at 3.6\,${\mu}$m in the S$^{4}$G catalogue. Their derived scalelength for the stellar disk is the same as ours, of $4$\,kpc. However, the bulge parameters are more similar to those from Merrit et al. \citeyear{2016ApJ...830...62M}. This shows that, irrespective of resolution, when using a self-consistent multiwavelength approach, with a more complex model for the inner regions, the result is very different. An inspection of the profiles from the S$^{4}$G catalogue and ours from Fig.~\ref{fig:M101_SB_profile2} shows very good fits in both cases, demonstrating a clear degeneracy, should only this band be used. We argue here the added value of the multiwavelength approach in going beyond the simple bulge/disk decomposition at individual wavelengths.

\subsection{SFR and sSFR for NGC~3938 and M101}
Our analysis indicates that both M101 and NGC~3938 are "active building" galaxies, consistent with their sSFR values. 
In this scheme, active building galaxies such as will eventually consume their gas and evolve to more quiescent galaxies with slightly lower star formation rates, and then eventually to massive spheroidal galaxies having consumed all their available gas for star formation (or by having had star formation halted otherwise).

The reciprocal of specific star formation rate, (sSFR)$^{-1}$, can be interpreted as the time over which a galaxy\textquotesingle s stellar mass has been built up assuming a constant rate of star formation. Thus a galaxy with a large sSFR can be thus interpreted as a  galaxy that has recently created a significant fraction of its stellar mass by consuming available ISM gas. The large sSFR value for M101, the increasing nature of the M101\textquotesingle s sSFR towards its outer radius (Fig.\ref{fig:N3938_M101_eta}), and the likely large reservoir of gas available to form stars suggest a picture of M101 as a spiral still in the process of vigorously forming stars. This view is supported by the fairly large fraction of gas mass for M101 compared to its total baryonic mass, M$_G$/(M$_\star$ + M$_G$), which we estimate as $0.75^{+0.18}_{-0.14}$ using our stellar mass estimate and taking the M101 gas mass as the sum of H and H$_2$ masses given by Aniano et al. \citeyear{2020ApJ...889..150A}. As a point of reference, in the MW gas comprises an estimated 13\% of the baryon mass (Kalberla \& Kerp \citeyear{2009ARA&A..47...27K}). The significantly increasing outward sSFR of M101 (in light of its probable interaction with companion galaxies seen in projection) supports a picture of star formation in the outer parts of M101 as a casual result of recent or current interactions with companions. 

Contrary to M101, the NGC 3938 sSFR radial distribution exhibits only a very modest increase outside the bulge region across the main disk (Fig. \ref{fig:N3938_M101_eta}). Additionally our model indicates a smaller SFR, a smaller sSFR, and a smaller gas mass fraction for NGC 3938 as opposed to M101, $0.50^{+0.16}_{-0.11}$, calculated as above for M101. NGC 3938 has more mass than M101 in the form of molecular hydrogen, 33\% as opposed to 18\% for M101 (Table 10, Aniano et al. \citeyear{2020ApJ...889..150A}). Unlike M101, NGC 3938 is not known to be currently interacting with nearby companions. When comparing these two galaxies, our model gives tentative to support the idea that NGC 3938 is a more typical quiescent star forming late type spiral than M101, and is further along in the process of converting existing  HI to H$_2$ and then to stars, with no obvious external influences to either amplify present day star formation or facilitate accretion of fresh quantities of HI through interactions with companions. 

\begin{figure}
    \includegraphics{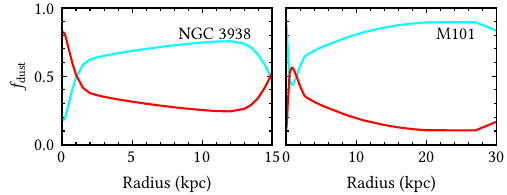}
    \caption{Our model prediction of the fraction of total energy absorbed by dust from the young (blue) and old (red) stellar populations is plotted as a function of distance from the galaxy centres for NGC 3938 and M101.}
    \label{fig:energy_abs2}
\end{figure}

\subsection{Radiative energetics}

The total model intrinsic stellar and dust luminosities of M101 and NGC 3938, broken down by constituent contribution, are given in Table \ref{tab:lumin_stellar}. Remarkably, despite notable differences between these two galaxies and their environments, the bolometric stellar and dust luminosities appear to be virtually identical.
However, our model predicts that the stellar disk and thin stellar disk  components contribute in different amounts to the total bolometric luminosity for M101 and NGC 3938. In NGC 3938 the stellar disk makes a larger contribution than the thin stellar disk, while in M101, both disks have similar contributions. This is in line with the previous findings, that M101 is a more active galaxy than NGC 3938, having an increased level of stellar luminosity coming from the young stars in the thin stellar disk.

The bulge contribution to the total stellar luminosity is 9.2\% of the total stellar luminosity for NGC 3938 and 1.2\% for M101.

 The fraction of the total luminosity of a galaxy absorbed (and re-emitted) by dust, which we denote as $f_{\rm{abs}}$
 is an important metric reflecting both the microscopic properties of dust grains, the overall dust mass, and distribution of dust in a galaxy relative to stellar heating sources.
 We can calculate $f_{\rm{abs}}$ from our model by dividing total galaxy dust luminosities from Table \ref{tab:lumin_stellar} by the intrinsic stellar luminosities from Table \ref{tab:lumin_stellar} obtaining values of 0.34 and 0.33 for NGC 3938 and M101, respectively. Bianchi \citeyear{2018A&A...620A.112B} in their study of 814 local galaxies of different morphological types from the DustPedia sample (Davies et al. \citeyear{2017PASP..129d4102D}) and modeled with the CIGALE found an average $f_{\rm{abs}}$ value of $\approx34.0\pm13.6\%$ from a sub-sample of Sc galaxies, in good agreement with our determinations.

 Our model allows us to ascertain the relative importance of dust heating contributed separately from the model thin stellar disk and stellar disk/bulge stars by calculating the fraction of the total energy emitted by dust that was absorbed through photons originating from a specific stellar population, $f_{\rm{dust}}$. For NGC 3938, the total contribution to
dust heating from thin stellar disk stars is $f_{\rm{dust}}^{\rm {young}}=72\%$, leaving a contribution of
$f_{\rm{dust}}^{\rm {old}}=28\%$ from stellar disk and bulge stars. For M101 we find heating contributions of $f_{\rm{dust}}^{\rm {young}}=78\%$  and $f_{\rm{dust}}^{\rm {old}}=22\%$. Our values of $f_{\rm{dust}}^{\rm {young}}$ are similar to those determined for the late type spirals  NGC 628, M33, NGC 891, MW of 71\%, $80$, 69\%, and 71\% , respectively, from application of the same RT model used here (Rushton et al. \citeyear{2022MNRAS.514..113R}; \citetalias{TP20}; \citetalias{PT11}; Natale et al. \citeyear{2022MNRAS.509.2339N}). Similar to the finding of Natale et al. \citeyear{2022MNRAS.509.2339N} for the MW, our $f_{\rm{dust}}^{\rm {young}}$ value also exceeds that found from the study of DustPedia archive of local galaxies by Nersesian et al. \citeyear{2019A&A...624A..80N}, who derived fractions from 52\%, 53\%, and 56\% (their quantity $S^{\rm abs}_{\rm young}$) for Hubble T stage 5, 6, and 7 galaxies corresponding approximately to Hubble type Sc. Natale et al. \citeyear{2022MNRAS.509.2339N} attribute this discrepancy to use of a simple energy balance model (CIGALE) by Nersesian et al. \citeyear{2019A&A...624A..80N} which ignores RT effects, as well as differences between our model and that of Nersesian et al. \citeyear{2019A&A...624A..80N} as to the definition of a young stellar population. Natale et al. \citeyear{2022MNRAS.509.2339N} note that RT modeling results of galaxies from other studies have tended to support fractional contributions to dust heating from young stars closer to our (and their) values (see Natale et al. \citeyear{2022MNRAS.509.2339N} discussion and references). Our results are thus consistent with the finding that the dominant contributor to dust heating in late-type spirals is from young stellar populations residing in a thin stellar disk. 

A detailed radial profile of $f_{\rm{dust}}$ may also be calculated from our models revealing interesting and nuanced details of dust heating which can be correlated with model stellar and dust component structures. Numerically $f_{\rm{dust}}$ as a function of galaxy central distance, $R$, is calculated by integrating absorbed stellar energy over the galaxy vertical extent at elliptical apertures from the galaxy centre. Radial distributions of $f_{\rm{dust}}$ are plotted in Fig. \ref{fig:energy_abs2} for NGC 3938 and M101 with the red curves tracing $f_{\rm{dust}}^{\rm {old}}$ due to emission from all stellar disk stars plus bulge stars, and the blue curves tracing $f_{\rm{dust}}^{\rm {young}}$  due to all thin stellar disk stars.

For NGC 3938, examining the left hand plots of Fig. \ref{fig:energy_abs2}, it is clearly evident, as with the total values discussed above, that the principle contribution to dust heating is from the model thin disk stellar population. Only at the centre the dust heating contribution due to old stars dominates, with $f_{\rm{dust}}^{\rm old}=0.8$, where most of the contribution is actually from bulge stars. Beyond the centre  $f_{\rm{dust}}^{\rm old}$ declines steeply to $R\approx0.7$ kpc, at which point $f_{\rm{dust}}^{\rm old}$ and $f_{\rm{dust}}^{\rm young }$ heating curves meet. For $R\gtrapprox0.7$ kpc $f_{\rm{dust}}^{\rm young }$ dominates the dust heating contribution.
Beyond $R\approx1$ kpc $f_{\rm{dust}}^{\rm old}$  declines and the contribution is due to stellar disk stars (as opposed to bulge stars), decreasing slowly and monotonically to $f_{\rm{dust}}^{\rm old}\approx0.27$. 

Radial distributions of $f_{\rm{dust}}$ for M101 are shown in the right-hand panel of Fig.\ref{fig:energy_abs2}. By contrast with NGC 3938, these distributions reflect the greater complexity of the M101 model. The behaviour of $f_{\rm{dust}}$ in the inner region of M101 is due to the combined effect of different overlapping stellar components extending to the galaxy centre. 
For dust in the inner 1 kpc of M101 we calculate that $f_{\rm{dust}}^{\rm young}=58$\%, with a remarkable 33\% being due to the nuclear thin disk alone. As noted in Sec. \ref{sec:SFR_surf_dens} above, the thin nuclear disk also accounts for the central singularity in star formation rate surface density. The compactness and intensity of this heated dust emission feature highlights the effectiveness of SB profiles made from thermal IR images at tracing the presence of compact central concentrations of UV emitting stars in galaxies, such nuclear disks of young stars, or compact massive star clusters.

For M101 generally, as with NGC 3938, the MW, and NGC 628, our RT model predicts the dust heating contribution from thin stellar disk populations tends to increase outwards.
Fig.\ref{fig:energy_abs2} indicates that past the inner, more complex region of the galaxy, for $R\gtrapprox0.5$ kpc, the heating contribution from thin stellar disk stars does indeed rise steadily, dominating the dust heating

\subsection{What we have learnt so far}

With the addition of the models for M101 and NGC~3938 we have now built a small sample of face-on nearby galaxies containing detailed information on intrinsic properties, with all galaxies modelled  using our radiative transfer codes (\citetalias{PT11}) and the same modelling technique and algorithms (\citetalias{TP20}). The sample consists of five face-on spirals: M33, NGC~628, M51, NGC~3938 and M101. To this we can add for reference the edge-on spiral Milky Way.

We note that relative comparisons between the intrinsic properties of our galaxies reduce the effects of systematic errors introduced when comparing the results of galaxies modeled using differing theoretical approaches. In addition systematic errors associated with data reduction approaches or adopted fitting procedures will  affect galaxies modelled the same way equally. Thus, in principle, differences found in derived physical parameters from our modelled galaxies will more likely be a direct result of actual physical differences between them.

A main question we wanted to address was to understand if face-on spirals, with their large variety of structure and spiral arms, can be fitted with axi-symmetric RT models over the whole range of UV-optical-NIR-FIR-submm data. With five galaxies in hand, we can now confirm that they can be successfully modelled in this way. The advantage of this type of RT modelling is that it allows the derivation of the large-scale geometry of the stellar and dust emissivity self-consistently with their luminosity. This type of modelling is the only one in existence for which the geometry is an output rather than an input. Non-axi-symmetric RT models for face-on galaxies do exist, but they need to input the geometry, effectively meaning they only fit the spatially integrated SEDs. They are thus not self-consistent and have, for this reason, less predictive power.

Having established the validity of our models, we can now try to understand what we can learn from the intrinsic properties we derived so far, and how we can extend this type of analysis to larger statistical samples.

The first thing to remark is that all the modelled galaxies were consistent with a MW type dust, with optical properties as in the models of \cite{2001ApJ...548..296W}. The only exception to this is M51, for which there is suggestion that a dust model with varying optical properties may provide a better fit to the PAH features and the 2200 \AA\ bump. \cite{2023MNRAS.526..118I} showed that a model with a LMC type dust in the inner disk of M51 could present an alternative solution. No extra constraints on this are available to give further consideration to this alternative fit. Overall we consider a MW type dust a good representation of the diffuse interstellar medium in nearby galaxies. Our analysis showed that a Milky Way type dust provides a self-consistent solution to the panchromatic SEDs of all galaxies in our sample. However, we cannot prove uniqueness in the solution.

Secondly, we do not find any submm excess in our fits. By contrary, we find a very good match to the submm emission of all galaxies, including M33, for which \cite{2010A&A...523A..20B,2016A&A...590A..56H} and \cite{2019MNRAS.487.2753W} suggested the existence of a submm excess.
We already commented in \citetalias{TP20} that the ability of our RT model to fit the so called "submm excess" of M33 without any need to modify the dust grain emissivity is due to the ability of our technique to fit the detailed geometry of the galaxy.

Thirdly, the panchromatic imaging analysis, coupled with the power of the radiative transfer calculations allowed us to disentangle various morphological components in our galaxies. Obviously this was also possible because all the galaxies were nearby, and had good panchromatic high resolution imaging. A main finding is that all galaxies seem to have a so called "main disk" with a characteristic inner radius, mainly revealed in the distribution of the young stars (our thin stellar disk), and in the dust distribution. In some cases this inner radius is also present in the distribution of the old stellar population (our stellar disk). For example in the Milky Way this inner radius was found at 4.5 kpc, which is the limit of the Milky Way bar. Beyond this radius all the galaxies show exponentially declining disks in both the stellar and dust distributions. Shortwards of this radius the galaxies show quite complex inner stellar structures like bars, rings and nuclear disks, and dust distributions that sometimes have central "holes". In our models this led to the introduction of linear functions to characterise the central stellar/dust emissivity declining towards the centre, with superimposed inner and nuclear disks and bulges to describe the complex behaviour of the central regions of galaxies. We consider thus the inner radius of the main disk of our galaxies an important internal characteristic, and we will use it as a reference point for comparing intrinsic distributions in galaxies of varying spatial extent.

In addition to the main disk we also found that some galaxies contained outer disks, which appear to be real distinct morphological components, as reflected in the distribution of their intrinsic properties. 

Our sample contains the following morphological components within the inner radius of the main disk:
\begin{itemize}
\item M33: nuclear disk, inner disk
\item NGC~628: inner disk, bulge
\item M51: inner disk, bulge
\item NGC~3938: inner disk, bulge
\item M101: nuclear disk, inner disk, bulge
\item MW: nuclear disk, inner disk, bulge
\end{itemize}
We should mention that the nuclear disk of the MW was not explicitly incorporated in our model, but left for further studies.

Beyond the main disk our sample has the following structure:
\begin{itemize}
\item M33: outer disk
\item NGC~628: -
\item M51 : outer disk
\item NGC~3938: -
\item M101: -
\item MW: -
\end{itemize}
We note that in the MW there is evidence for an outer disk, but the imaging data on some wavebands was not deep enough to properly allow the study of an outer disk.

\begin{figure}
	\includegraphics[width=8cm]{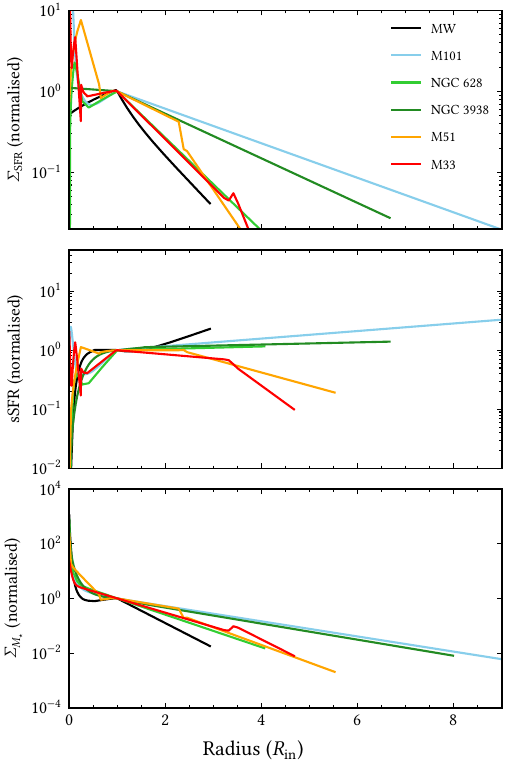}
    \caption{Comparison of key stellar quantities derived as a function of radius for every star forming spiral galaxy modeled using our RT model. In the  above panels, from top to bottom, are plotted: star formation rate surface density $\Sigma_{\mbox{SFR}}$,  specific star formation rate $\mbox{s}\mbox{SFR}$, and stellar mass surface density $\Sigma_{\mbox{M}^{\star}}$. To facilitate comparison of results between different galaxies, radial distances
    are normalised to that of the inner radius of each galaxy\textquotesingle s thin stellar disk (for galaxies modeled with more than one thin stellar disk normalisation is to the inner radius of the galaxy\textquotesingle s main thin stellar disk). Magnitudes of the plotted quantities are normalised to their thin stellar disk inner radii values, thus all plotted values are dimensionless. The fit results of this paper were used to create the M101 and NGC 3938 plots and those of Natale et al. \citeyear{2022MNRAS.509.2339N}, Rushton et al. \citeyear{2022MNRAS.514..113R}, Inman et al. \citeyear{2023MNRAS.526..118I} and \citetalias{TP20} for the plots for MW, NGC 628, M51, and M33, respectively.}
 
    \label{fig:gal_comparison}
\end{figure}

\subsubsection{The radial distribution of key stellar population parameters}

An important feature of the RT transfer model employed for this paper is its ability to model the properties of galactic stellar populations by fitting observational images while properly accounting for significant dust absorption and scattering effects. By using images as opposed to only SEDs to fit the model, the model gives information on the spatial distribution of stellar properties (within the limitations of our assumption of axisymmetry). As was noted in \citetalias{PT11}, another key feature of our RT model is its ability to fit intrinsic stellar galaxy SEDs without the need for a priori assumptions of stellar population synthesis. Modeling approaches based, for instance, solely on the assumption of energy balance, lack these key features. To this end we find it useful to analyse the radial distributions of key stellar population parameters,  namely SFR surface density, stellar mass surface density, and specific SFR, for all the galaxies in our small sample.

In Fig. \ref{fig:gal_comparison} we have plotted the star formation rate surface density $\Sigma_{\mbox{SFR}}$,  specific star formation rate $\mbox{s}\mbox{SFR}$ , and stellar mass surface density $\Sigma_{\mbox{M}^{\ast}}$ for our sample galaxies as a function of centre distance $R$. As mentioned before, we make use of this important structural characteristic found for our galaxies, the inner radius of the main (thin) disk, to better enable comparisons between our galaxies of varying scale. We have thus normalised all $R$ values to the galaxy main thin stellar disk inner radius and the plotted quantities themselves are normalised to their values at this radius. 

The SFR surface densities reveal primarily exponential decline in the main disk of all galaxies, with a wide range of scale lengths, the longest M101, the shortest the MW. 

The sSFR values beyond the inner radii show a more or less constant behaviour within their main disk (NGC~3938, NGC~628, M51), as expected for spiral galaxies, or a slight decline (M33). Exceptions to this is are the MW and M101. In particular M101 has an increase of a factor of $\approx5$. This extreme increase may indicate star formation associated with interacting star forming companions. Overall it shows a main disk that is vigorously growing at the present epoch, with an inside-out formation scenario still actively shaping the galaxy at the present epoch.

Stellar mass surface density values, $\Sigma_{\mbox{M}^{\star}}$, seen in the bottom panel of Fig. \ref{fig:gal_comparison} all decline exponentially outwards. Interestingly the decline rates of M101, NGC 3938, and M51 are nearly identical and are the least of our small sample. 
As with $\Sigma_{\mbox{SFR}}$ and sSFR,  $\Sigma_{\mbox{M}^{\ast}}$ for M33 shows significant change of slope starting at its outer disk ($R/R_{in}\gtrsim3.3$). This behaviour provides support for the physical existence of an outer disk as a basic structural component of M33 with likely variant epoch of formation and evolution from its inner components. 
 \begin{figure}
    \centering
    \includegraphics[width=\columnwidth]{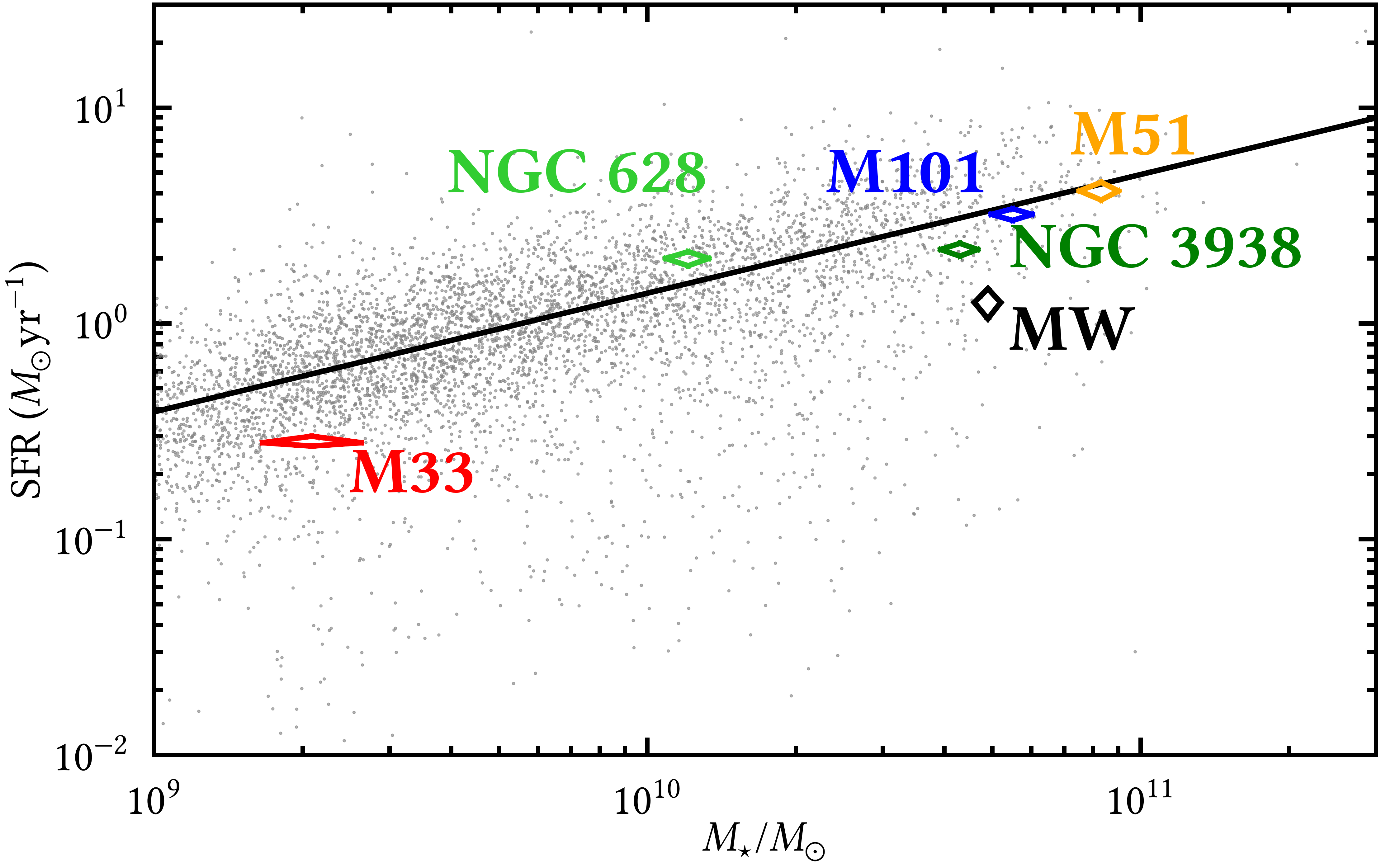}   
    \caption{SFR versus stellar mass $M_\star$ for our sample of nearby galaxies: M33, NGC 628, NGC 3938, M101, M51 and the MW (rhomboid symbols). The main sequence of galaxies is defined by a reference sample (grey dots) of local Universe disk-dominated galaxies taken from \protect\cite{2017AJ....153..111G}. The vertical and horizontal apexes of the rhomboid symbols are placed at the $1\sigma$ bounds in SFR and $M_\star$, respectively. The solid line is the regression fit to a single power-law model given in table 2 of \protect\cite{2018MNRAS.477.1015G}}
    \label{fig:sfr-mass}
\end{figure}

 \subsubsection{A structurally resolved relation for spiral galaxies}

It is well known that the star-formation rate and the stellar mass of star forming galaxies follow a relation called the main sequence (MS) of galaxies (e.g. \cite{2007ApJ...660L..43N, 2015A&A...575A..74S, 2017A&A...608A..41C}), with deviations from the sequence interpreted in the context of stellar/ISM evolutionary and AGN effects. We expect all our modelled galaxies to follow this relation. Indeed, looking at Fig.~\ref{fig:sfr-mass}, the 6 galaxies follow the trend defined by a reference sample, chosen from \cite{2017AJ....153..111G}. The reference sample is a flux-limited subset (complete to 19.4 mag in the the SDSS r band) of 5202 disc- dominated GAMA (Galaxy And Mass Assembly Survey; \cite{Driver2011}, galaxies located within redshift 0.13, and which are not classified as belonging to a galaxy group (see Table 1 and section 4.3 of \cite{2017AJ....153..111G}).

While NGC 628, M101, M51 follow closely the MS, as defined by the regression fit to data from the reference sample, M33, MW, and NGC~3938 are slightly below the fit, though still within the scatter of the relation. This means that M33, the MW and NGC~3938 are overall more quiescent spirals, a result already noted in \citetalias{TP20} for M33 and \cite{2022MNRAS.509.2339N} and the MW, and in the discussion above for NGC~3938. In this scheme, active building galaxies such as NGC~628, M101, and M51  will eventually consume their gas and evolve to more quiescent galaxies with slightly lower star formation rates for equal stellar mass (like the MW).   

\begin{figure}
    \centering
    \includegraphics[width=\columnwidth]{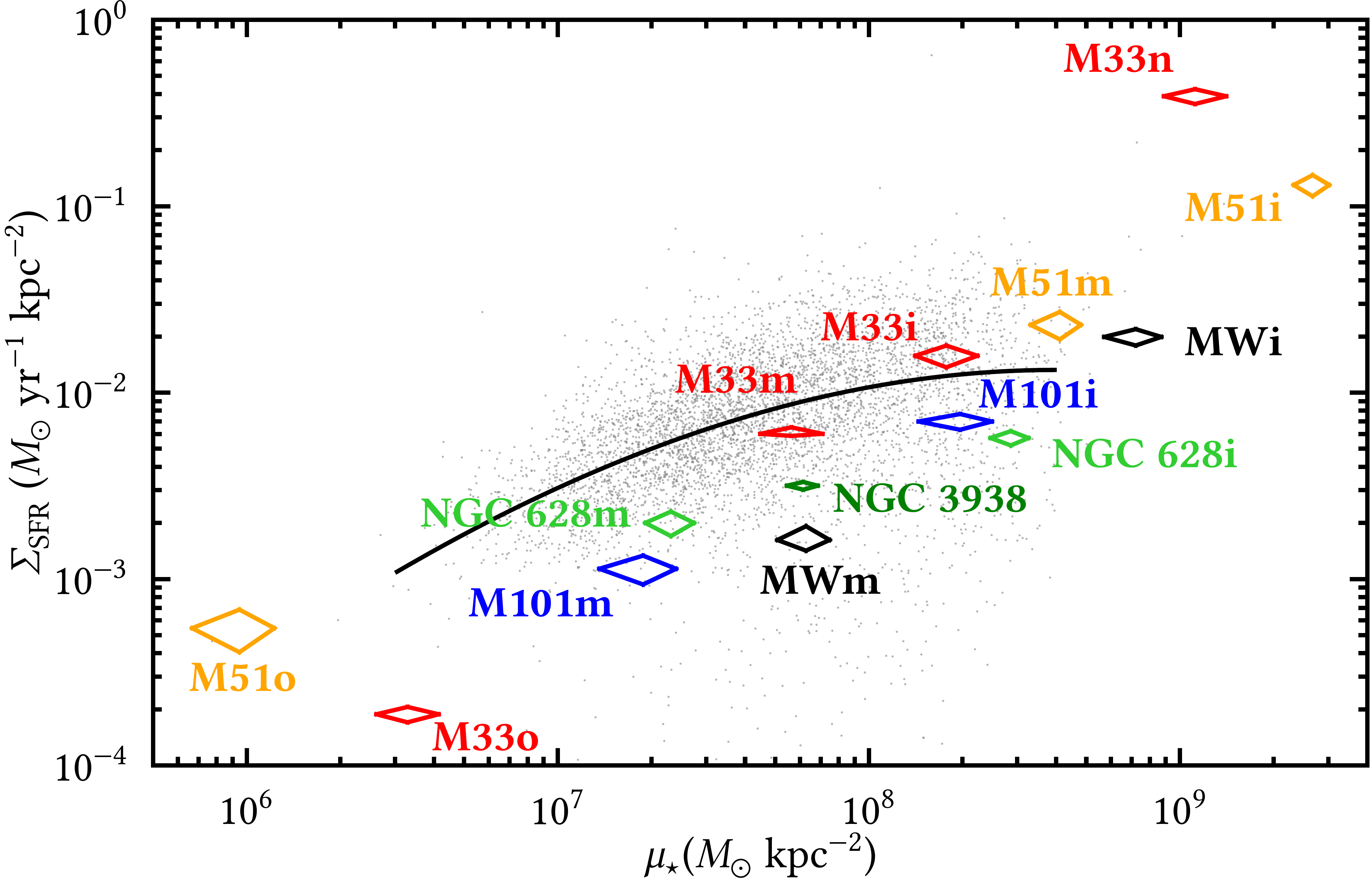}   
    \caption{Right: The SFR surface density $\Sigma_{\rm SFR}$ versus stellar mass surface density ($\mu_\star$) for the morphological components of our sample of nearby galaxies. Symbols and reference sample as in the left panel. The different morphological components are labelled using the standard notation from this paper. For example for M33 we have M33n - the nuclear disk; M33i - the inner disk; M33m - the main disk; M33o - the outer disk.}
    \label{fig:sfr-density}
\end{figure}

Having derived the SFR and the $M_\star$ for each of the morphological components of the analysed galaxies, we can define a MS based on surface densities,  $\Sigma_{\rm SFR}$ and stellar mass surface density ($\mu_{*}$). For the GAMA galaxies in the reference sample surface densities were calculated within the $R_{\rm eff}$ in the intrinsic i band, with $R_{\rm eff}$ derived from S\'ersic surface photometry. The normalised (to the surface area) MS relation of the \cite{2017AJ....153..111G} sample is plotted in Fig.~\ref{fig:sfr-density}. Overplotted we show the "structurally resolved relation" (SRR, as introduced in \citetalias{TP20}) for the individual morphological components of our galaxies. We can see that for NGC~628 the inner disk - main disk SRR follows a track parallel with the MS. NGC~3938 does not have inner and outer disks. However, all the other galaxies show SRRs that follow a much steeper relation, with outer disks departing from the MS in the green valley direction, and nuclear disks displaced towards star-burst regions. The steepest and most extreme relation is found for M33 followed by M51. The MW also has a very steep relation. Taken into account that the outer disk and nuclear disk of the MW have not been decomposed in our analysis, probably an extended SRR for the MW would look as steep as that for M33. 

The new SRRs show that the different morphological components have suffered different evolutionary paths, or have been affected differently by feedback and environmental mechanisms. They almost look like different galaxies glued together, with the well behaved NGC~628 being the exception rather than the rule. We already discussed in \citetalias{TP20} that a possible explanation for the behaviour of the SRR is related to the fact that stellar feedback can operate on a variety of spatial and temporal scales. 
This, in turn, induces a variation in SF activity on the corresponding spatial and temporal scales, which may lead the SFR cycle to be out of phase between the different morphological components, nuclear disks, inner disks, main disks and outer disks.

\section{CONCLUSIONS}
\label{sec:conclusions}

We used the radiative transfer model of \citetalias{PT11} and the optimisation technique from \citetalias{TP20} to model the nearby face-on galaxies M101 and NGC~3938. For both galaxies the model provides excellent fits to the azimuthally-averaged surface-brightness profiles over the whole range of UV-optical-FIR-submm wavelengths. We derived the large-scale distribution of stars and dust, and corresponding fundamental intrinsic properties: SFR, sSFR, stellar mass, dust opacity, dust mass and dust attenuation. We found the following:

\begin{itemize}
\item
${\rm SFR=3.2\pm0.2}$ ${\rm M}_{\odot}$\,yr$^{-1}$ and ${\rm SFR=2.2\pm0.1}$ ${\rm M}_{\odot}$\,yr$^{-1}$ for M101 and NGC~3938, respectively.
\item
The average sSFR values are $5.8^{+2.5}_{-1.3}\times10^{-11}$ yr$^{-1}$ and $5.0^{+0.57}_{-0.31}\times10^{-11}$ yr$^{-1}$ for M101 and NGC 3938, respectively.
\item The fraction of the total luminosity absorbed by dust, $f_{\rm abs}$, is 0.34 and 0.33 for NGC~3938 and M101, respectively.
\item The contribution to dust heating from the young stellar population is $72 \%$ and $78\%$ for NGC~3938 and M101, respectively.
\item
The sSFR of NGC~3938 is approximately constant throughout most of its extent, with only a very small increase towards larger radii. By contrast, the sSFR (of the main disk) of M101 shows a large increase, by a factor of five.
This could be due to recent interactions with companions.
Overall, our model  supports a picture whereby NGC~3938 is a more typical quiescent spiral than M101. It also supports the idea that NGC~3938 is further along in the process of converting existing HI to H$_2$ and then to stars, than M101, with no obvious external influences to either amplify present day star formation or facilitate accretion of fresh quantities of HI through interactions with companions. 

\end{itemize}

With the addition of the models for M101 and NGC~3938 we built a small sample of face-on nearby galaxies containing detailed information on intrinsic properties, with all galaxies modelled using our radiative transfer codes \citepalias{PT11} and the same modelling technique
and algorithms \citepalias{TP20}. The sample consists of five face-on spirals:
M33, NGC 628, M51, NGC 3938 and M101. To this we added for
reference the edge-on spiral Milky Way.
The decoding of the spectral energy distribution of these six galaxies allowed us to draw the following conclusions:
\begin{itemize}
\item Axi-symmetric RT codes can successfully account for the spectral and spatial distribution of face-on spirals, over the whole spectral range from the UV to the FIR and submm.
\item Milky Way-type dust with optical properties from \cite{2001ApJ...548..296W} provides a self-consistent solution to
the panchromatic SEDs of all galaxies of our sample.
\item We do not find any submm excess in our fits. We attribute the good match to the observed submm emission to the ability of our modelling technique to fit the detailed geometry of a galaxy.
\item Dust absorbs around $35\%$ of the total stellar luminosity, with a scatter in the percentage fraction of around $3-4\%$.
\item The young stellar populations dominate the dust heating, contributing on average to $74\%$, with a scatter in the percentage fraction of around $5-6\%$.
\item The panchromatic imaging modelling allows us to disentangle various morphological components: bulges, nuclear disks, inner disks, main disks and outer disks.  A main finding is that all
galaxies seem to have a so called "main disk" with a characteristic
inner radius, mainly revealed in the distribution of the young stars and in the dust distribution. Shortwards of
this radius the galaxies show quite complex inner stellar and dust structures.
\item The sSFR beyond the inner radii show a more or less constant value within their main disk. A notable exception is M101, for which the sSFR increases by a factor of $\approx5$ from the inner to the truncation radius. It shows a main disk that is vigorously growing at the present epoch.
\item NGC~628, M101 and  M51 follow closely the main sequence (MS) relation for star-forming galaxies. M33, the MW and NGC~3938 are slightly below the fit to the relation, but still within the observed scatter.
\item In the SFR surface density versus stellar mass surface density space we find a structural resolved relation (SRR) for the morphological components of our galaxies, that is steeper than the main sequence. The exception to this is for NGC~628, where the SRR is parallel to the MS. The new SRRs show that the different morphological components have suffered different evolutionary paths, or have been affected differently by feedback and environmental mechanisms.
    \end{itemize}

\section*{ACKNOWLEDGEMENTS}
We would like to thank an anonymous referee for a careful reading of the manuscript and for insightful comments that helped improve the paper. Christopher J. Inman acknowledges support from a Science and Technology Facilities Council studentship grant (grant number ST/W507386/1). 

This work is based in part on observations made with the NASA Galaxy Evolution Explorer. GALEX is operated for NASA by the California Institute of Technology under NASA contract NAS5-98034.  
This research has made use of the NASA/IPAC Infrared Science Archive, which is operated by the Jet Propulsion Laboratory, California Institute of Technology, under contract with the National Aeronautics and Space Administration.

This work is also based on the Sloan Digital Sky Survey (SDSS) data.
Funding for the SDSS IV has been provided by the Alfred P. Sloan Foundation, the U.S. Department of Energy Office of Science, and the Participating Institutions. SDSS acknowledges support and resources from the Center for High-Performance Computing at the University of Utah. The SDSS web site is www.sdss4.org.
SDSS is managed by the Astrophysical Research Consortium for the Participating Institutions of the SDSS Collaboration including the Brazilian Participation Group, the Carnegie Institution for Science, Carnegie Mellon University, Center for Astrophysics | Harvard \& Smithsonian (CfA), the Chilean Participation Group, the French Participation Group, Instituto de Astrofísica de Canarias, The Johns Hopkins University, Kavli Institute for the Physics and Mathematics of the Universe (IPMU) / University of Tokyo, the Korean Participation Group, Lawrence Berkeley National Laboratory, Leibniz Institut für Astrophysik Potsdam (AIP), Max-Planck-Institut für Astronomie (MPIA Heidelberg), Max-Planck-Institut für Astrophysik (MPA Garching), Max-Planck-Institut für Extraterrestrische Physik (MPE), National Astronomical Observatories of China, New Mexico State University, New York University, University of Notre Dame, Observatório Nacional / MCTI, The Ohio State University, Pennsylvania State University, Shanghai Astronomical Observatory, United Kingdom Participation Group, Universidad Nacional Autónoma de México, University of Arizona, University of Colorado Boulder, University of Oxford, University of Portsmouth, University of Utah, University of Virginia, University of Washington, University of Wisconsin, Vanderbilt University, and Yale University.

This work has also made use of data products from the Two Micron All Sky Survey, which is a joint project of the University of Massachusetts and the Infrared Processing and Analysis Center/California Institute of Technology, funded by the National Aeronautics and Space Administration and the National Science Foundation. 
This work is based in part on observations made with the {\it Spitzer} Space Telescope, which is operated by the Jet Propulsion Laboratory, California Institute of Technology under a contract with NASA.
We also utilise observations performed with the ESA {\it Herschel} Space Observatory \citep{2010A&A...518L...1P}, in particular to do photometry using the PACS  \citep{2010A&A...518L...2P} and SPIRE \citep{2010A&A...518L...3G} instruments.

\section*{DATA AVAILABILITY}
The data underlying this article will be shared on reasonable request to the corresponding author.


\bibliographystyle{mnras}
\bibliography{./References.bib}

\newpage
\appendix
\onecolumn
\section{Stellar emissivity profiles}
\FloatBarrier
\begin{figure}
    \centering
    \includegraphics[width=6cm]{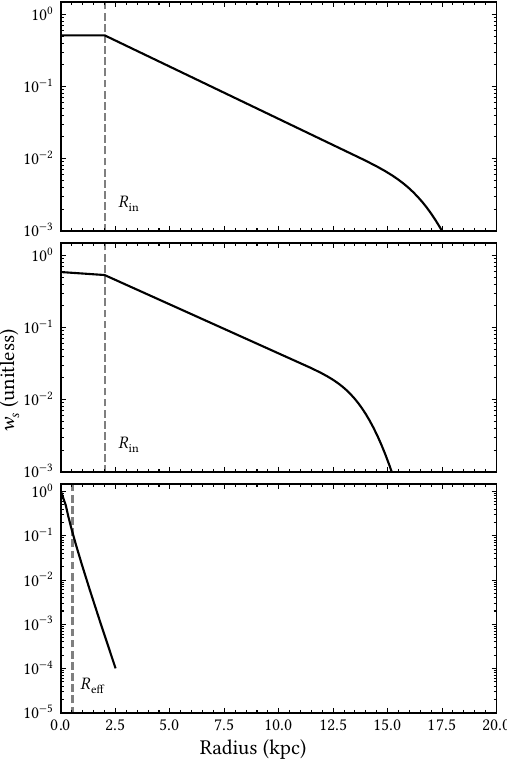}
    \caption{Examples of model stellar emissivity radial distributions in the SDSS g-band for the disk, thin disk and bulge at $z=0$. The distributions correspond to the best fit model of NGC~3938. Plotted as a vertical dashed line is the position of the inner radius for the disk and thin disk, and the effective radius for the bulge.}
\label{fig:radial_distribution}
\end{figure}

\newpage
\section{The effect of different sampling in photometry}
\label{app:photometry}

\begin{figure}
    \centering
\includegraphics[width=10cm]{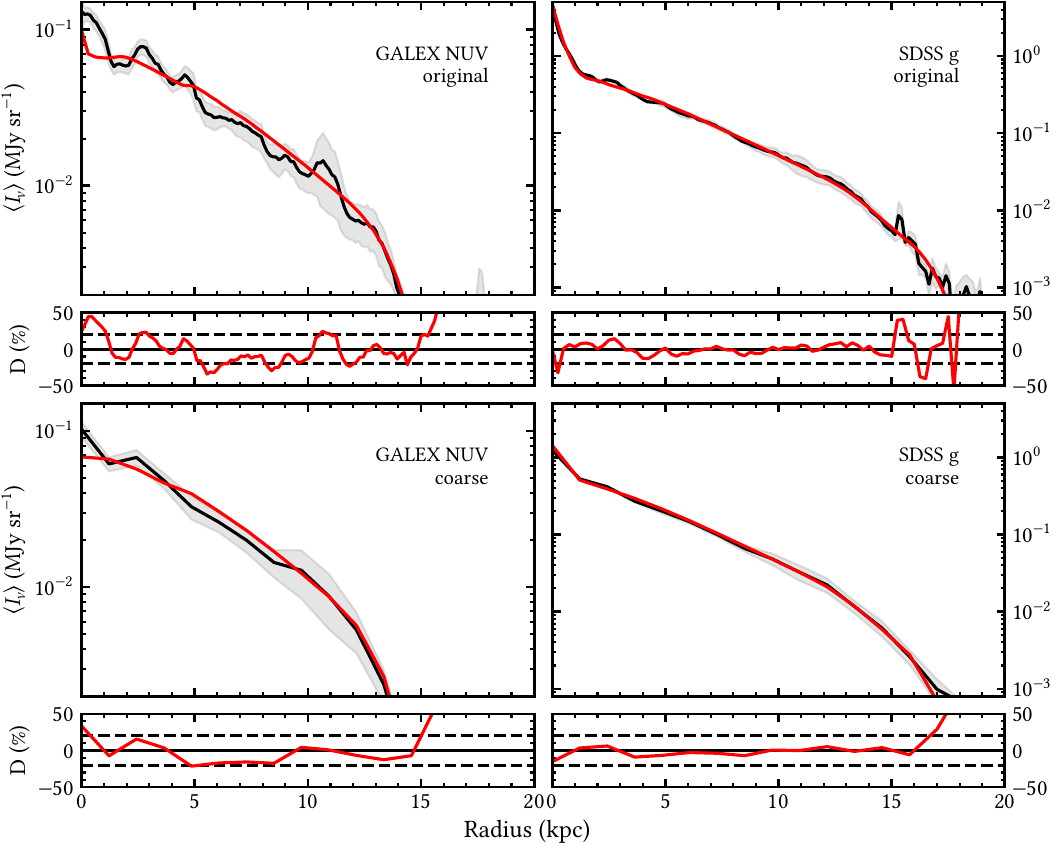}   
    \caption{Azimuthally averaged surface-brightness profiles produced with a fixed 2-pixel size aperture (top) and with a fixed physical sampling (of the lowest resolution data - the Herschel 500\,${\mu}$m). The black line shows the data and the red line the model. The left panels show the GALEX NUV profiles while the right panels show the SDSS g-band profiles. The percent differences between the best fit model SB values and the observed SB values, $D_{\lambda_l n}$[\%] (see Sec. $\ref{sec:fitqual}$), are plotted in the sub-plot below each profile, with the dotted black horizontal lines indicating +20\%, and -20\% deviation.}  
    \label{fig:aperture_test}
\end{figure}

In Sect.~\ref{sec:SB_prof} we discussed that the photometry adopted in this paper is done by using a fixed 2-pixel aperture. This means that the surface-brightness profiles sample different physical scales at different wavelengths. We stressed the importance of using this procedure to get the best constraints on the model. In this appendix we show the effect of using an aperture that samples a fixed physical scale. For this we consider the physical scale sampled by our lowest resolution data, the Herschel 500\,${\mu}m$. We applied this sampling to the best fitted model images (fitted at the resolution of the original data) and to the corresponding 
observed images. Examples of this procedure are shown for NGC~3938 in Fig.~\ref{fig:aperture_test}, for the GALEX NUV and the SDSS g-band.

We note that the overall characteristics of the profiles remain the same, but the sharp central increase in the g-band profile due to the bulge is strongly smoothed. Although the best fit model seems to fit well the smoother profiles, it would be more difficult to do the fit and obtain a reliable bulge-disk decomposition in this low sampling mode. Otherwise a visual inspection of the profiles identifies the same truncations and inner radii for the model. 

The calculated residuals decrease due to the smoother character of the data, but the reduced ${\rm chi}^2$ of the fit is slightly increased.
\FloatBarrier

\newpage
\section{M101 AND NGC 3938 parameters}
\FloatBarrier
\begin{table}
    \centering
    \caption{Fixed geometrical parameters for the NGC 3938 RT model. All scale length and distance parameters have units of kpc. }
    \begin{tabular}{ | ll | c }
        \hline
        \multicolumn{3}{c}{}    \\ [-1ex]
        \multicolumn{3}{c}{Fixed Parameters} \\[-2ex]
        \multicolumn{3}{c}{}    \\
        \hline
        from data\\
        \hline
        disk inner truncation &      $R_{\rm tin, s}^{\rm disk}$     & 0.0    \\
        thin disk inner truncation   &  $R_{\rm tin, s}^{\rm tdisk}$    & 0.0   \\
        dust disk inner truncation  &      $R_{\rm tin, d}^{\rm disk}$     & 0.1   \\
        disk inner radius  &    $R_{\rm in, s}^{\rm disk}$      & $2.0$  \\
        thin disk inner radius  &    $R_{\rm in, s}^{\rm tdisk}$     & $2.0$ \\
        dust disk inner radius   &    $R_{\rm in, d}^{\rm disk}$      & $2.0$ \\

        disk truncation radius &   $R_{\rm t,s}^{\rm disk}$     & 17.0 \\
        thin disk truncation radius  &     $R_{\rm t,s}^{\rm tdisk}$    & 14.0 \\
        dust disk truncation radius &    $R_{\rm t,d}^{\rm disk}$     & 14.0 \\
        thin dust disk truncation radius &   $R_{\rm t,d}^{\rm tdisk}$    & 14.0 \\
        
        bulge S\'ersic index &  $n_{\rm s}$     & 1 \\
        bulge axial ratio & $b/a$ \, {\rm (S\'ersic)}   & 0.94
        
        \\
        \hline
        from model\\
        \hline
        thin dust disk scalelength & $h_{\rm d}^{\rm tdisk}$         & 3.2 \\
        thin dust disk inner radius & $R_{\rm in, d}^{\rm tdisk}$     & 2.0 \\
        thin dust disk inner truncation radius & $R_{\rm tin, d}^{\rm tdisk}$    & 0.0 \\   
        disk scaleheight & $z_{\rm s}^{\rm disk}$  & 0.19 \\
        thin disk scaleheight & $z_{\rm s}^{\rm tdisk}$ & 0.09 \\
        dust disk scaleheight & $z_{\rm d}^{\rm disk}$  & 0.16 \\
        thin dust disk scaleheight & $z_{\rm d}^{\rm tdisk}$ & 0.09 \\
        \hline
    \end{tabular}
    \label{tab:fixed_n3938} 
\end{table}

\begin{table*}
    \caption{Fixed geometrical parameters for the M101 RT model. All scale length and distance parameters have units of kpc. }
    \begin{tabular}{ | ll | c }
        \hline
        \multicolumn{3}{c}{}    \\ [-1ex]
        \multicolumn{3}{c}{Fixed Parameters} \\[-2ex]
        \multicolumn{3}{c}{}    \\
        \hline
        from data\\
        \hline
        
        inner and main disk inner truncation &     $R_{\rm tin,s}^{(\rm i-disk,m-disk)} $              & $(0, 0)$ \\
        nuclear, inner and main thin disk inner truncation &     $R_{\rm tin,s}^{(\rm n-tdisk,i-tdisk,m-tdisk)}$     & $(0, 0, 0)$ \\
        inner and main dust disk inner truncation &   $R_{\rm tin,d}^{(\rm i-disk,m-disk)}$               & $(0, 0)$ \\
        inner and main disk inner radius $R_{\rm in,s}^{(\rm i-disk,m-disk)} $               & $(0, 0)$ \\
        nuclear, inner and main thin disk inner radius &  $R_{\rm in,s}^{(\rm n-tdisk,i-tdisk,m-tdisk)}$      & $(0, 0, 2.5)$ \\
        inner and main dust disk inner radius & $R_{\rm in,d}^{(\rm i-disk,m-disk)}$                & $(0, 2.5)$ \\
        \\
        
        inner and main disk truncation radius & $R_{\rm t,s}^{(\rm i-disk,m-disk)}$             & $(5.0, 30.0)$ \\
        nuclear, inner and main thin disk truncation radius & $R_{\rm t,s}^{(\rm n-tdisk,i-tdisk,m-tdisk)}$   & $(1.0, 5.0, 30.0)$ \\
        inner and main dust disk truncation radius & $R_{\rm t,d}^{(\rm i-disk,m-disk)}$             & $(5.0, 30.0)$ \\
        \\
        bulge S\'ersic index & $n_{\rm s}$ & 2 \\
        bulge axial ratio & $b/a$ \, {\rm (S\'ersic)} & 0.6\\
        \hline
        from model\\
        \hline
        inner thin dust disk scalelength & $h_{\rm d}^{\rm i-tdisk}$   & $0.40\pm 0.04$ \\
        main thin dust disk scalelength & $h_{\rm d}^{\rm m-tdisk}$   & $5.2\pm 0.54$ \\
        \\
        inner and main thin dust disk inner truncation & $R_{\rm tin,d}^{(\rm i-tdisk,m-tdisk)}$     & $(0, 0, 1.029)$ \\
        inner and main thin dust disk truncation & $R_{\rm t,d}^{(\rm i-tdisk,m-tdisk)}$       & $(1.0, 5.0, 30.0)$ \\
        \\
        disk scaleheight & $z_{\rm s}^{\rm disk}$  & 0.40 \\
        dust disk scaleheight & $z_{\rm d}^{\rm disk}$  & 0.27 \\
        thin disk scaleheight & $z_{\rm s}^{\rm tdisk}$ & 0.09 \\
        thin dust disk scaleheight & $z_{\rm d}^{\rm tdisk}$ & 0.09 \\
        \hline
    \end{tabular}
    \label{tab:fixed_M101} 
\end{table*}

\begin{table*}
    \centering
	\caption{Model intrinsic stellar luminosity densities of NGC 3938 in W Hz$^{-1}$ of the bulge and stellar disks at each observed wavelength.}
	\label{tab:intr_lumin_N3938}
    \begin{tabular}{cccc}
        \hline
        $\lambda (\mu$m) & $L^{\rm bulge}_\nu$ & $L^{\rm disk}_\nu$ & $L^{\rm tdisk}_\nu$ \\ 
        \hline
        0.150 & - & - & 1.512$\times 10^{21}$ \\  
        0.220 & - & - & 2.679$\times 10^{21}$ \\ 
        0.365 & 1.848$\times 10^{20}$ & 1.159$\times 10^{21}$ & 2.426$\times 10^{21}$ \\ 
        0.443 & 9.479$\times 10^{20}$ & 5.266$\times 10^{21}$ & 3.804$\times 10^{21}$ \\ 
        0.564 & 1.511$\times 10^{21}$ & 8.082$\times 10^{21}$ & 3.949$\times 10^{21}$ \\ 
        0.809 & 2.059$\times 10^{21}$ & 1.208$\times 10^{22}$ & 2.415$\times 10^{21}$ \\ 
        1.259 & 4.026$\times 10^{21}$ & 2.365$\times 10^{22}$ & 3.948$\times 10^{20}$ \\ 
        2.200 & 3.302$\times 10^{21}$ & 2.000$\times 10^{22}$ & 3.097$\times 10^{20}$ \\ 
        3.600 & 1.450$\times 10^{21}$ & 1.068$\times 10^{22}$ & 6.028$\times 10^{19}$ \\ 
        4.500 & 9.509$\times 10^{20}$ & 6.673$\times 10^{21}$ & 6.028$\times 10^{19}$ \\ 
        5.800 & 6.381$\times 10^{20}$ & 1.487$\times 10^{22}$ & 6.028$\times 10^{19}$ \\
        \hline
    \end{tabular}
\end{table*}

\begin{table*}
\centering
\caption{Model intrinsic stellar luminosity densities of M101 in W Hz$^{-1}$ of the bulge and all stellar disks at each observed wavelength.}
\label{tab:intr_lumin_M101}
\begin{tabular}{ccccccc} 
    \hline
    $\lambda (\mu$m) & $L^{\rm bulge}_\nu$ & $L^{\rm i-disk}_\nu$ & $L^{\rm m-disk}_\nu$ & $L^{\rm n-tdisk}_\nu$ & $L^{\rm i-tdisk}_\nu$ & $L^{\rm m-tdisk}_\nu$ \\
    \hline
        0.150 & - & - & - & 5.294$\times10^{19}$ & 1.578$\times10^{19}$ & 2.533$\times10^{21}$ \\ 
        0.220 & - & - & - & 8.940$\times10^{19}$ & 7.401$\times10^{19}$ & 3.774$\times10^{21}$ \\ 
        0.365 & 6.886$\times10^{19}$ & 1.986$\times10^{19}$ & 1.369$\times10^{21}$ & 4.302$\times10^{19}$ & 5.828$\times10^{19}$ & 3.748$\times10^{21}$ \\ 
        0.480 & 1.653$\times10^{20}$ & 6.952$\times10^{19}$ & 6.540$\times10^{21}$ & 3.023$\times10^{19}$ & 1.001$\times10^{20}$ & 4.963$\times10^{21}$ \\ 
        0.620 & 3.250$\times10^{20}$ & 1.562$\times10^{20}$ & 1.009$\times10^{22}$ & 2.197$\times10^{19}$ & 1.746$\times10^{20}$ & 6.496$\times10^{21}$ \\ 
        0.770 & 1.354$\times10^{20}$ & 2.034$\times10^{20}$ & 1.128$\times10^{22}$ & 3.358$\times10^{19}$ & 2.668$\times10^{20}$ & 6.805$\times10^{21}$ \\ 
        1.260 & 5.839$\times10^{20}$ & 5.260$\times10^{20}$ & 2.033$\times10^{22}$ & 1.098$\times10^{19}$ & 2.182$\times10^{19}$ & 1.236$\times10^{21}$ \\ 
        2.200 & 4.879$\times10^{20}$ & 4.733$\times10^{20}$ & 2.064$\times10^{22}$ & 8.613$\times10^{18}$ & 1.711$\times10^{19}$ & 9.697$\times10^{20}$ \\ 
        3.600 & 3.933$\times10^{20}$ & 8.032$\times10^{18}$ & 1.394$\times10^{22}$ & 1.676$\times10^{18}$ & 3.331$\times10^{18}$ & 1.888$\times10^{20}$ \\ 
        4.500 & 1.452$\times10^{20}$ & 1.227$\times10^{20}$ & 8.583$\times10^{21}$ & 1.676$\times10^{18}$ & 3.331$\times10^{18}$ & 1.862$\times10^{20}$ \\ 
        5.670 & 7.988$\times10^{19}$ & 1.314$\times10^{20}$ & 1.688$\times10^{22}$ & 1.676$\times10^{18}$ & 3.331$\times10^{18}$ & 1.862$\times10^{20}$ \\
\hline
\end{tabular}
\end{table*}

\begin{table*}
\caption{This table gives the fit quality for the best fit model SB profiles to the observed SB profiles for NGC 3938 and M101 SB at selected  fitted wavebands. Column 1 lists the observed waveband name. Columns 2-5 list reduced chi square estimates and average percent deviations between the best model fit profiles from the data profiles at selected fitted wavebands, for each galaxy (Eq.\textquotesingle s $\ref{eq:Ravg}$ - $\ref{eq:chilam}$).}
\begin{tabular}{ l | c | c | c | c }
    \hline
    & \multicolumn{2}{ c }{NGC 3938}  &  \multicolumn{2}{c}{M101}\\
    
    \cmidrule(lr){2-3}\cmidrule(lr){4-5}
    
    Filter & ${\rm chi}_{\lambda_l}^2 $ & $ \overline{D_{\lambda_l}}[\%]$ & ${\rm chi}_{\lambda_l}^2 $ & $\overline{D_{\lambda_l}}[\%]$ \\
    \hline 
    GALEX NUV & 4.9   & 16.8  & 0.70  & 20.5          \\
    SDSS u    & 3.4   & 14.1  & 1.02  & 15.2          \\
    SDSS g    & 4.0   & 9.2   & 0.49  & 8.45          \\
    SDSS r    & 5.4   & 11.5  & 0.62  & 9.61           \\
    SDSS i    & 4.2   & 9.6   & 0.69  & 9.03           \\
    2MASS J   & 0.3   & 12.1  & 0.15  & 11.8           \\
    2MASS K   & 0.2   & 7.9   & 0.13  & 14.7           \\
    IRAC 1    & 4.6   & 14.4  & 1.65  & 9.1          \\
    IRAC 2    & 2.6   & 12.7  & 1.51  & 8.9          \\
    IRAC 3    & 8.8   & 14.5  & 5.36  & 22.4         \\
    SPIRE 500 & 3.7   & 16.6  & 0.59  & 10.9      \\
    \hline
\end{tabular}
\label{tab:fit_quality}
\end{table*}

\newpage


\bsp	
\label{lastpage}
\end{document}